\documentclass{aa}
\usepackage{multirow}
\usepackage{comment}

\usepackage{amsmath}
\usepackage{graphicx,subfig} 
\usepackage{lscape}
\usepackage{longtable}
\usepackage{txfonts}
\usepackage{natbib}
\usepackage{color}
\usepackage{booktabs}
\usepackage{siunitx}
\usepackage{comment}
\usepackage[dvipsnames]{xcolor}
\usepackage{dblfloatfix}
\usepackage{enumitem}

\makeatletter
\renewcommand*\aa@pageof{, page \thepage{} of \pageref*{LastPage}}
\usepackage{hyperref}
\hypersetup{colorlinks=true,allcolors=[rgb]{0,0,0.8}}

\renewcommand{\arraystretch}{1.2}
\newcommand{\erosita}{eROSITA\xspace}

\newcommand{\extlike}{\ensuremath{\mathcal{L}_\mathrm{ext}}\xspace}

\newcommand{\diff}{\mathrm{d}}

\defcitealias{Ghirardini2024}{G24}

\newcommand{\ncattot}{200,000\xspace}
\newcommand{\ramcattot}{125~GB\xspace}
\newcommand{\nepochtrain}{20,000\xspace}
\newcommand{\nepochval}{10,000\xspace}
\newcommand{\regsubsample}{20,000\xspace}
\newcommand{\epochsredraw}{20\xspace}
\newcommand{\regintermediatedim}{30\xspace}
\newcommand{\reglearningrate}{\ensuremath{1 \times 10^{-5}}\xspace}
\newcommand{\regbs}{100\xspace}
\newcommand{\regconverge}{125\xspace}

\newcommand{\regtraintime}{\ensuremath{\SI{2}{\hour}}\xspace}

\newcommand{\sbilearningrate}{\ensuremath{3 \times 10^{-5}}\xspace}
\newcommand{\sbibs}{256\xspace}
\newcommand{\sbiconverge}{40\xspace}

\newcommand{\sbimades}{8\xspace}
\newcommand{\sbitransforms}{256\xspace}

\newcommand{\rsqtrain}{0.9340}
\newcommand{\rsqval}{0.9237}
\newcommand{\rsqtrainomegam}{0.9153}
\newcommand{\rsqvalomegam}{0.8987}
\newcommand{\rsqtrainsigmaeight}{0.9274}
\newcommand{\rsqvalsigmaeight}{0.9178}

\newcommand{\deltaomegam}{0.033\xspace}
\newcommand{\deltasigmaeight}{0.037\xspace}
\newcommand{\deltaAx}{0.192\xspace}
\newcommand{\deltaBx}{0.099\xspace}
\newcommand{\deltaFx}{0.280\xspace}
\newcommand{\deltaGx}{0.322\xspace}
\newcommand{\deltasigmax}{0.235\xspace}

\newcommand{\percentdeltaomegam}{11.5\xspace} % 0.033/0.29 * 100
\newcommand{\percentdeltasigmaeight}{4.4\xspace} % 0.037/0.85 * 100

\DeclareSIUnit \parsec {pc}

\usepackage{natbib,twoopt}
\usepackage{xcolor}
\defcitealias{Ghirardini2024}{G24}
\makeatletter
\newcommandtwoopt{\citeads}[3][][]{\href{http://adsabs.harvard.edu/abs/#3}%
    {\def\hyper@linkstart##1##2{}%
     \let\hyper@linkend\@empty\citealp[#1][#2]{#3}}}
  \newcommandtwoopt{\citepads}[3][][]{\href{http://adsabs.harvard.edu/abs/#3}%
    {\def\hyper@linkstart##1##2{}%
     \let\hyper@linkend\@empty\citep[#1][#2]{#3}}}
  \newcommandtwoopt{\citetads}[3][][]{\href{http://adsabs.harvard.edu/abs/#3}%
    {\def\hyper@linkstart##1##2{}%
     \let\hyper@linkend\@empty\citet[#1][#2]{#3}}}
  \newcommandtwoopt{\citeyearads}[3][][]%
    {\href{http://adsabs.harvard.edu/abs/#3}
    {\def\hyper@linkstart##1##2{}%
     \let\hyper@linkend\@empty\citeyear[#1][#2]{#3}}}
\makeatother

\title{Simulation-Based Inference for Cluster Cosmology with Set-Based Neural Network Architectures}

\author{S.~Zelmer\inst{1}, % szelmer@mpe.mpg.de
E.~Bulbul\inst{1,2}, % ebulbul@mpe.mpg.de
K.~Lehman\inst{3,4,5},
S.~Krippendorf\inst{6}, 
E.~Artis\inst{1}, % eartis@mpe.mpg.de
S.~Grandis\inst{7},
N.~Clerc\inst{8}, % nicolas.clerc@irap.omp.eu
Z.~Ding\inst{1},
L.~Fiorino\inst{1},
V.~Ghirardini\inst{9},
M.~Kluge\inst{1},
N.~Malavasi\inst{1},
A.~Merloni\inst{1},
T.~Mistele\inst{1},
K.~Nandra\inst{1},
M.~E.~Ramos-Ceja\inst{1},
J.~S.~Sanders\inst{1},
F.~Pacaud\inst{10},
A.~von~der~Linden\inst{11},
J.~Weller\inst{1,3,5},
and X.~Zhang\inst{1}
} 

\institute{
Max-Planck-Institut f\"ur extraterrestrische Physik, Giessenbachstrasse 1, 85748 Garching, Germany %1
\and
Fakult\"at f\"ur Physik, LMU M\"unchen, Schellingstr. 4, 80799 München, Germany %2
\and
Universit\"ats-Sternwarte M\"unchen, Fakult\"at f\"ur Physik, LMU M\"unchen, Scheinerstr. 1, 81679 M\"unchen, Germany %3
\and
Center for Computational Astrophysics, Flatiron Institute, 162 5th Avenue, New York, NY, 10010, USA %4
\and
Excellence Cluster ORIGINS, Boltzmannstr. 2, 85748 Garching, Germany %5
\and
University of Cambridge, DAMTP and Cavendish Laboratory, Cambridge CB3 0WA, United Kingdom %6
\and
Universit\"at Innsbruck, Institut f\"ur Astro- und Teilchenphysik, Technikerstr. 25/8, 6020 Innsbruck, Austria %7
\and
Univ Toulouse, CNES, CNRS, IRAP, Toulouse, France %8
\and 
INAF, Osservatorio di Astrofisica e Scienza dello Spazio, via Piero Gobetti 93/3, I-40129 Bologna, Italy %9
\and
Argelander-Institut f\"ur Astronomie, Universität Bonn, Auf dem H\"ugel 71, 53121 Bonn, Germany %10
\and
Department of Physics and Astronomy, Stony Brook University, Stony Brook, NY 11794, USA %11
}
\date{\today}

\titlerunning{Cluster Cosmology using Simulation-based Inference with Set-based Neural Networks}
\authorrunning{Zelmer et al.}

\begin{document}
\abstract{The unprecedented statistical power of galaxy cluster catalogs from the SRG (Spectrum Roentgen Gamma)/eROSITA All-Sky Survey provides a unique opportunity to place stringent constraints on cosmological models through measurements of structure growth. Fully exploiting the potential of these large X-ray-selected cluster samples, however, requires robust statistical frameworks that accurately connect observable quantities to the underlying cosmological parameters.
We develop and implement a simulation-based inference (SBI) framework for cosmological parameter estimation using a realistic mock-generation pipeline calibrated on eRASS1 simulations. Synthetic galaxy cluster catalogs are propagated through the survey selection function to produce mock eRASS1 observations that reproduce the data's statistical properties. At the core of the method lies a set-based neural network (i.e., a graph neural network operating on sets) that encodes information from individual clusters and is coupled to a masked autoregressive flow for flexible posterior density estimation. This approach enables the use of the full cluster-level information content without compressing the observables into binned summary statistics.
Our framework recovers the input cosmologies within the inferred uncertainties, and passes simulation-based calibration tests (SBC) and TARP diagnostics, demonstrating robustness in the presence of realistic survey effects, including selection biases and intrinsic scatter in the scaling relations. We obtain mock constraints of \percentdeltaomegam\% on $\Omega_\mathrm{m}$ and \percentdeltasigmaeight\% on $\sigma_8$ averaged over a suite of simulated cluster catalogs matching the effective sample size of the data set (of order 3,300 clusters). We achieve a precision comparable to that obtained with traditional MCMC analyses based on substantially larger cluster samples. The framework is readily extensible to more complex forward models and additional observables. This work highlights the potential of SBI methods for next-generation large-scale structure analyses with forthcoming X-ray cluster surveys.
}

\keywords{galaxies: clusters: general --
galaxies: clusters: intracluster medium --
(cosmology:) cosmological parameters --
cosmology: observations --
(cosmology:) large-scale structure of the Universe -- methods: statistical}

\maketitle

\section{Introduction}

Galaxy cluster 
%are the most massive gravitationally bound structures in the Universe and are an established probe for cosmology. Their 
abundances as a function of mass and redshift encode the growth of cosmic structure and are particularly sensitive to the matter density parameter, $\Omega_{\mathrm{m}}$, and the amplitude of matter fluctuations, often expressed as $\sigma_{8}$ or a combination of both, $S_{8}$. Large, homogeneously selected samples of clusters therefore provide a powerful avenue for constraining cosmology and testing models of structure formation (see, e.g., \citealt{Abbott2025}, \citealt{Lesci2025}, \citealt{Miyatake2023}, \citealt{Costanzi2021}, \citealt{Artis2025}, \citealt{Ghirardini2024} (\citetalias{Ghirardini2024} hereafter), \citealt{Garrel2022}, \citealt{Aymerich2025}, \citealt{Bocquet2024}).

The Spectrum-Roentgen-Gamma (SRG) mission, with its soft X-ray instrument eROSITA (extended ROentgen Survey with an Imaging Telescope Array), conducted the first imaging all-sky X-ray survey since ROSAT \citep{Voges1999}, and with unprecedented depth \citep{Sunyaev2021, Predehl2021}. During the operation phase, eROSITA performed approximately 4.4 All-Sky Surveys between 2019 and 2022. Already, the first all-sky survey (eRASS1) has delivered a catalog containing 5,259 extent-selected galaxy clusters in the Western Galactic Hemisphere \citep{Bulbul2024, Kluge2024}. This sample is complemented by the clusters in the point-source sample \cite{Balzer2025}, which represents the largest ICM-selected cluster sample currently available. A multitude of cosmological studies have been performed based on this catalog, including a main cosmological analysis \citepalias{Ghirardini2024}, investigations of the growth of structure and general relativity \citep{Artis2025}, alternative $f(R)$ gravity models \citep{Artis2024}, constraints on ultra-light axion dark matter \citep{Zelmer2025}, and cluster clustering analyses \citep{Seppi2024}. Furthermore, machine learning algorithms have previously been applied to eROSITA data to infer the masses of galaxy groups and clusters without assuming an underlying dark matter profile or hydrostatic equilibrium, and have been shown, based on hydrodynamical simulations, to accurately recover halo masses down to $10^{13}$~M$_\odot$ based on the hydrodynamical simulations \citep{Krippendorf2024}.

Extracting cosmological constraints from such datasets requires accurate modeling of the survey selection function and high accuracy of observable--mass relations and their scatter, while fully exploiting the statistical power of the catalog \citep{Grandis2024}. Standard analyses have relied on Markov Chain Monte Carlo (MCMC) methods, which typically involve the numerical computation of the expected number density of observed clusters with a likelihood function (e.g. \citealt{Bocquet2019}; \citetalias{Ghirardini2024}; \citealt{Costanzi2019}; \citealt{Boehringer2014}).
These approaches are primarily limited by the model complexity. Modeling effects include correlations among multiple observables, skewness in the distributions, complex scaling relations, and sample variance, which typically lead to a large number of nuisance parameters that must be integrated out numerically. Furthermore, the expansion in sample size, together with the increased volume of ancillary weak-lensing data employed in mass calibration, results in greater computational demands and, consequently, longer likelihood evaluation times. Finally, they usually involve an information loss inherent to data compression.

A promising alternative is simulation-based inference (SBI; see \citealt{Cranmer2020} for a review), a likelihood-free approach that infers posterior distributions of model parameters directly from mock observations generated under forward simulations. In the context of cluster cosmology, SBI enables the use of the full information contained in individual clusters, while naturally incorporating survey selection effects and complex scaling relations. Neural density estimators, such as normalizing flows (e.g., masked autoregressive flows \citep{Papamakarios2017}), offer a flexible way of approximating the posterior, provided suitable data embeddings are used. Since mock catalogs from cosmological forward models vary in the number of objects they contain, the corresponding SBI architecture must be able to process inputs of varying sizes. Consequently, some operation needs to be applied to the variable-dimensional data vector to standardize its dimensionality; the easiest choice is binning in the data space. 

In this work, we take a step further and present the first application of SBI to Bayesian cosmological inference for X-ray-selected galaxy clusters that does not require fixed input vector dimensions and can handle varying sample sizes. Using mock catalogs calibrated to the eRASS1 survey, we train a set-based embedding network that encodes individual cluster properties and passes them to a masked autoregressive flow for posterior estimation. The set-based neural network resembles a graph neural network (GNN) that operates on sets (i.e., graphs without edges). This allows us to use highly informative neural-informed summary statistics rather than naive compression techniques such as binning. We demonstrate, using realistic mock catalogs calibrated to the eRASS1 survey, that this method can recover Bayesian cosmological constraints consistent with established approaches, while providing a framework capable of scaling to future eROSITA data releases. This work focuses on methodological development and validation, while applications to observational survey data are deferred to follow-up work.

This paper is structured as follows. Section~\ref{sect:trainingdata} describes the simulated mock catalogs used in this work. Section~\ref{sect:architecture} introduces our SBI framework, including the embedding network and density estimator. In Sect.~\ref{sect:results}, we present the results of our cosmological inference and discuss the implications and robustness of our findings. Section~\ref{sect:conclusion} provides our conclusions and outlook. 

\section{Training data}
\label{sect:trainingdata}

To train a neural network to infer cosmological and auxiliary parameters from eRASS1, we adopt a forward model that closely mirrors the survey (see Artis et al., in prep., 2026). The full parameter set used in \citetalias{Ghirardini2024} consists of 23 free parameters (five cosmological parameters, five X-ray and optical scaling relation parameters, four gravitational weak lensing mass bias correction parameters, two correlation coefficients, and two contamination fraction parameters). The convergence of current SBI methods appears to be harder for higher dimensions. Hence, we simplify the model by focusing on a well-motivated subset of eleven parameters: all cosmological parameters, the scaling relation parameters relating the X-ray observable to mass, and the correlation between the X-ray observable and weak lensing mass. A summary of all parameters used in this work as well as the ones used by \citetalias{Ghirardini2024} is listed in Tab.~\ref{tab:paramlist}. To further reduce the number of parameters in the sample construction, we make a modeling choice of removing contaminant-related parameters. We compensate for this by adopting the higher-purity sample used in previous eROSITA cosmology analyses. Specifically, we apply a strict extension-likelihood ($\extlike>10$), resulting in a sample purity of 97\% \citep{Bulbul2024}.

We obtain mock catalogs following four steps that are described in detail below:

\begin{enumerate}[leftmargin=1em, itemindent=1em]
    \item Sample from the galaxy cluster halo mass function (HMF)
    \item Apply scaling relations and intrinsic (correlated) scatter
    \item Perform selection and apply instrument scatter
    \item Model tangential shear profiles (where applicable)
\end{enumerate}

\begin{figure}[h!]
    \centering
    \includegraphics[width=0.48\textwidth]{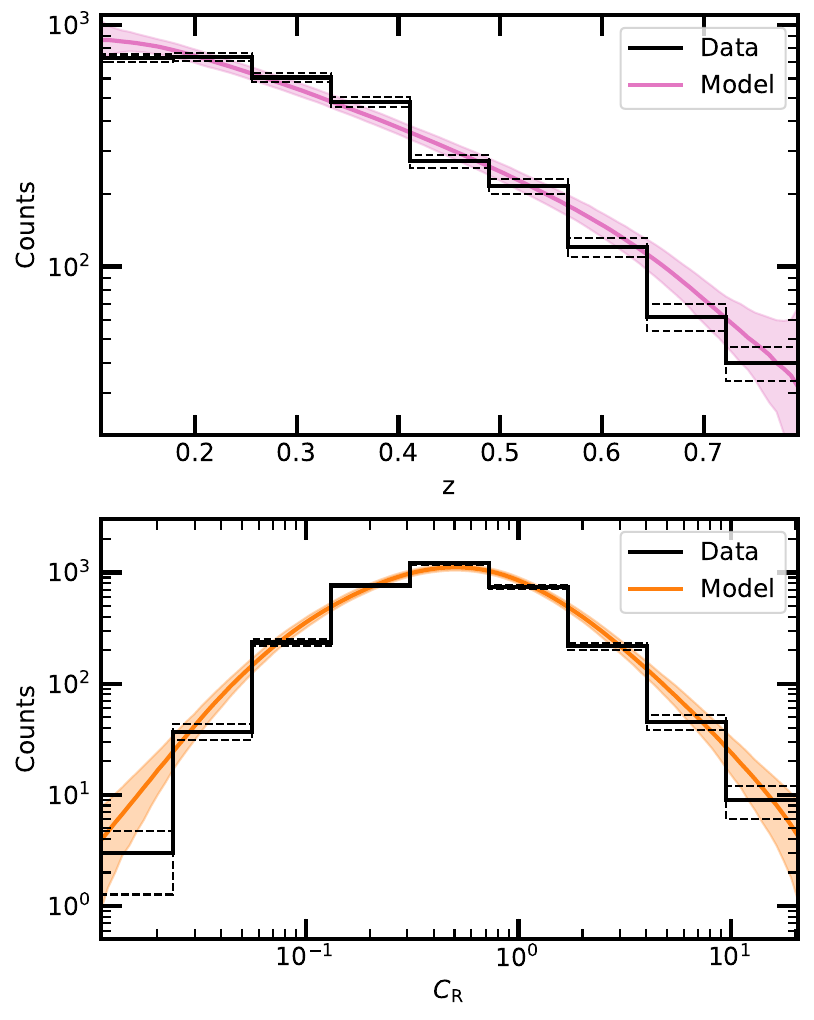}
    \caption{Plot of the binned eRASS1 catalog with $\extlike>10$ in nine redshift (top panel) and logarithmic count-rate (bottom panel) bins in black against the best-fit model (pink and orange). The best-fit parameters are obtained from mock realizations based on posterior draws from a likelihood-based posterior estimation (MCMC) on the eRASS1 data. The errors on the data histogram and the model lines correspond to 1$\sigma$ uncertainties.}
    \label{fig:model}
\end{figure}

Additional simplifying assumptions are introduced in the modeling to ensure the pipeline remains tractable within the SBI framework, such as omitting the uncertainties of the weak-lensing mass bias. Furthermore, the optical scaling relation is omitted on the basis that it is expected to contribute limited cosmological constraining power and primarily serves to exclude low-richness contaminant systems from the sample. Since the high-extent likelihood cutoff ensures that contaminants are negligible, this assumption does not introduce a significant bias into our analysis. Figure~\ref{fig:model} shows that the chosen model can describe the data well.
The last column of Tab. \ref{tab:paramlist} indicates parameters used in the forward model for the training data. The reason behind choosing this subset of parameters for this work is explained in this section. The details on the choice of priors are given in Sect.~\ref{sect:performance}.

\subsection{Sampling from the halo mass function}
\label{sect:hmfsampling}
The HMF returns the mass and redshift distributions of dark matter halos for a given set of cosmological parameters. Following the conventions of \citetalias{Ghirardini2024}, we refer to the mass $M_{500c}$ of a halo by the inclosed mass in a sphere with an average density that is 500 times larger than the critical energy density of the Universe, $\rho_\mathrm{crit}(z)=3H(z)^2/(8\pi G)$, at redshift $z=0$. Here, $H(z)$ is the Hubble rate and $G$ is Newton's gravitational constant. The comoving number density of halos $n$ per unit logarithmic mass is
\begin{equation}
    \frac{\diff n}{\diff \ln M}(M, z) = f(\sigma)\,\frac{\bar{\rho}_m}{M}\,\frac{\diff \ln\sigma^{-1}}{\diff \ln M},
\end{equation}
\noindent where $\bar{\rho}_m$ is the mean matter density at $z=0$, $f$ is called the multiplicity function and is typically fitted by simulations \citep[here, we use the fitting function obtained by][]{Tinker2008}, and $\sigma$ is the variance of the matter power spectrum defined as
\begin{equation}
    \sigma^{2}(M, z)
    = \frac{1}{2\pi^{2}}
      \int_{0}^{\infty} k^{2} P(k; z)\,\tilde{W}^{2}(kR)\, dk,
\end{equation}
\noindent with radius $R = 3M/(4\pi \bar{\rho}_m)^{1/3}$ and the top-hat window function $\tilde{W}(x)=3(\sin(x)/x^3-\cos(x)/x^2)$.

Once a fitting function, a top-hat function, and a mass convention are chosen, the HMF depends solely on the cosmological parameters. In this work, we vary the following set of cosmological parameters: The matter density fraction of the Universe, $\Omega_\mathrm{m}$, the variance of the matter power spectrum at $R=8~\si{\mega\parsec}/h$, $\sigma_8$, the Hubble constant in \si{\kilo\meter\per\second\per\mega\parsec}, $H_0$, the baryon matter density fraction of the Universe, $\Omega_\mathrm{b}$, and the scalar spectral index, $n_\mathrm{s}$. The other free parameters of the matter power spectrum are fixed to the \cite{Planck2020} best-fit values, as these parameters are not sensitive to galaxy cluster number counts.

The resulting HMF is a density function in two variables (logarithmic mass $\log(M)$ and redshift $z$). The mock samples are drawn using rejection sampling on the 2D grid in the regions $0.01<z<0.9$ (101 bins) and $13.1<\log(M/$M$_\odot)<16$ (91 bins). The optically confirmed eRASS1 cosmology sample clusters cover the redshift range $0.1 < z < 0.8$ by construction. The mass range has been chosen to encompass a broad interval around the estimated physical masses of the eRASS1 galaxy clusters. The lower limit should be set as low as possible to properly account for the Eddington bias but as high as necessary to keep computational time manageable. We checked that any lower mass limit below $\log(M_\mathrm{min})\approx13.5$ does not bias the results after selection. We thus chose a conservative lower limit of $\log(M_\mathrm{min})~\approx~13.1$ and only sampled dark matter halos above this threshold.

In this step, we typically sample $5~\times~10^7$ dark matter haloes in the eRASS1 survey footprint (for a cosmology around the best fit parameters by \citetalias{Ghirardini2024}) in a time $t_\mathrm{DM}$ of $t_\mathrm{DM}\approx6$~s. The time effort varies widely depending on the choice of input parameters, since the number of clusters in the catalog is sensitive to the HMF associated to the input cosmology as well as the scaling parameters in the selection process.

\subsection{Scaling relations and intrinsic scatter}

This subsection describes how the intrinsic properties, i.e., the true count-rates and weak-lensing (WL) masses, are obtained from the intrinsic halo masses and redshifts. The necessary steps include applying a mass-X-ray scaling relation, a mass-WL-mass scaling relation based on the weak lensing mass bias, and computing the intrinsic (and correlated) scatter of the intrinsic quantities.

We emphasize that cluster richness is not produced within this framework, as optical information is primarily used to ensure high sample purity in the standard eROSITA cosmology set-up. Incorporating richness estimates would require introducing a full six-parameter scaling relation, which is beyond the capabilities of our current SBI framework. We compensate for this by applying a higher $\extlike>10$ cut compared to a lower $\extlike>6$ of \citetalias{Ghirardini2024}. 
The intrinsic quantities (count-rate $C_\mathrm{R}$ and WL mass $M_\mathrm{WL}$ are computed within a time $t_\mathrm{int}$ with $t_\mathrm{int} \approx 3$~s (for the best fit parameters by \citetalias{Ghirardini2024}). The computation of the correlated intrinsic scatter constitutes the most computationally expensive part of this forward modeling step.

\subsubsection{X-ray scaling relation}

Under the assumption of self-similarity, galaxy clusters are expected to follow power-law relations (scaling relations) between their masses and observables \citep{Kaiser1986}. The approach has been shown to successfully model cluster populations if gauged on gravitational weak lensing masses of the subset of clusters with available weak lensing shear data \citep{vonderLinden2014, Chiu2022, Robertson2024}. We employ the scaling relation formalism in the mock generator by computing a weak lensing mass given a fixed weak lensing mass bias and assuming a power law between cluster mass $M$ and the X-ray observable count-rate $C_\mathrm{R}$, defined as the number of counts from the object divided by the exposure time:

\begin{equation}
\label{eq:xray_scaling}
\begin{split}
    \left\langle \ln \frac{C_\mathrm{R}}{C_\mathrm{R,p}} \Bigg| M, z \right\rangle =\,& A_\mathrm{X} + B_\mathrm{X}\ln\frac{M}{M_\mathrm{p}} + D_\mathrm{X}\ln\frac{d_\mathrm{L}(z)}{d_\mathrm{L}(z_\mathrm{p})} + E_\mathrm{X}\ln\frac{E(z)}{E(z_\mathrm{p})} \\
    &+ F_\mathrm{X}\ln\frac{M}{M_\mathrm{p}}\ln\frac{1+z}{1+z_\mathrm{p}} + G_\mathrm{X}\ln\frac{1+z}{1+z_\mathrm{p}},
\end{split}
\end{equation}

\noindent where $A_{X}$ and $B_{X}$ are the normalization and slope. We choose to fix $D_\mathrm{X}=-2$ and $E_\mathrm{X}=2$ following the self-similar model, commonly used in cluster scaling relations \citep{Pratt2019, Bulbul2019, Bahar2022, Ramos-Ceja2025}, while keeping the redshift evolution terms $F_\mathrm{X}$ and $G_\mathrm{X}$ free to allow for more flexibility in the scaling law. Equation \ref{eq:xray_scaling} is used to associate the mass and redshift of a cluster to an idealized count-rate. The count-rates obtained using this scaling law will receive an intrinsic scatter $\sigma_\mathrm{X}$ correlated to the scatter of the weak lensing masses.

\subsubsection{Weak lensing mass calibration}

A highly accurate mass proxy of a galaxy cluster can be inferred from the lensed background galaxies behind that cluster by fitting a density profile to the cluster \citep[e.g.][]{Grandis2024, Giocoli2024} or by applying direct methods \citep{Mistele2024, Mistele2025}. In the former case, these WL masses have been found to be biased (WL mass bias) due to astrophysical effects and miscentering with respect to the intrinsic mass of a cluster \citep[e.g.][]{Becker2011, Bahe2012, Rasia2012, Sommer2025}.

For the training data, the WL masses are obtained from the intrinsic cluster masses following a scaling relation based on the WL mass bias \citep{Grandis2021}:

\begin{equation}
\label{eq:wl_scaling}
    \left\langle \ln \frac{M_\mathrm{WL}}{M_\mathrm{WL,p}} \Bigg| M, z \right\rangle = b(z) + b_\mathrm{M}\ln\frac{M}{M_\mathrm{p}}.
\end{equation}

\noindent Here, $b(z)$ is the survey-dependent redshift evolution of the WL mass bias and $b_\mathrm{M}$ the mass trend of the WL mass bias. A similar model with survey-dependent redshift evolution $s(z)$ and a mass trend $s_\mathrm{M}$ is assumed for the intrinsic scatter of the WL masses:

\begin{equation}
    \ln \sigma_\mathrm{WL}^2 = s(z) + s_\mathrm{M} \ln \frac{M}{M_\mathrm{p}}.
\end{equation}

\noindent The intrinsic WL mass scatter $\sigma_\mathrm{WL}$ is assumed to be correlated with the intrinsic count-rate scatter $\sigma_\mathrm{X}$ with a correlation coefficient $\rho_\mathrm{X,WL}$ that is left free to vary in the mock simulations \citep{Grandis2024}.

Among the weak lensing surveys used for the standard eRASS1 cosmological analysis, the Dark Energy Survey (DES) \citep{Abbott2022, Grandis2024}, KiDS \citep{Kleinebreil2025}, and HSC \citep{Okabe2025, Chiu2025} weak lensing measurements have been utilized for mass calibration, while DES has the largest joint footprint with eRASS1 and thus the greatest statistical power. For the SBI pipeline, for simplicity, we model only DES-like WL masses and shear profiles, given the significantly greater signal-to-noise ratio and impact on the overall results compared to KiDS and HSC. 
%The details of the modeling of the DES WL masses are explained in \cite{Grandis2024, Ghirardini2024}. 
In the light of an extensive sky coverage of the eRASS1 footprint provided by the combined DECADE \citep{Anbajagane2025} and DES weak lensing shear catalogs (Mistele et al., in prep., 2026), we constructed shear profiles for every cluster within our SBI training data.

\subsection{Selection and instrument scatter}

The galaxy cluster selection is modeled using the eRASS1 count-rate selection function, including the $\extlike>10$ cut \citep{Clerc2024}. The selection function is trained on an eRASS1's digital twin, with a full light cone generated by N-body simulations and baryon painting, using the published scaling relations in the literature \citep{Comparat2020, Seppi2022}. 

Before applying the selection function, the observed redshift $\hat{z}$ is precomputed from the intrinsic redshift $z$. All objects with observed redshift smaller than the maximum eRASS1 survey depth at that specific position are removed \citep[see,][for further details]{Kluge2024}. Additionally, only clusters inside the eRASS1 cosmology range $0.1 \leq \hat{z} \leq 0.8$ are kept. These cuts reduce the number of objects entering the selection function by approximately 50\%, resulting in a factor 2 increase in runtime performance. The observed count-rate is computed using the eROSITA instrument-specific scatter. For the best-fit parameters of eRASS1 cosmology, approximately $3\times10^3$ clusters are selected, and the selection process takes time $t_\mathrm{sel}$ with $t_\mathrm{sel} \approx 7$~s due to the large input catalog with intrinsic quantities of $\mathcal{O}(10^7)$ clusters (see Sect.~\ref{sect:hmfsampling}).

\subsection{Tangential shear profiles}
\label{sect:shears}

The tangential reduced shear profile as a function of angular distance $g_\mathrm{t}(\theta)$ is computed from the intrinsic quantities (WL mass $M_\mathrm{WL}$ and observed redshift $\hat{z}$ of the lens) as well as the DES source distribution:

\begin{equation}
    g_\mathrm{t}(\mathbf{\theta}; M_\mathrm{WL}, \hat{z}) = \frac{\gamma(\mathbf{\theta}; M_\mathrm{WL}, \hat{z})}{1 - \kappa(\mathbf{\theta}; M_\mathrm{WL}, \hat{z})}.
\end{equation}
\noindent Here, $\gamma(\theta)$ and $\kappa(\theta)$ are the tangential components of the shear and the convergence, respectively. We follow the modeling of $\gamma(\theta)$ and $\kappa(\theta)$ \citep{Grandis2024}. 

Typically, there is a miscentering between the true gravitaional centers and the X-ray centers that are used for the WL analysis. To model this effect, a conditional variational auto-encoder (CVAE) is trained to resample the miscentering distribution based on the eRASS1 digital twin by \cite{Seppi2022}. The details of this auxiliary neural network are given in Appendix \ref{app:cvae}.

The tangential reduced shear profiles are evaluated at six redshift-dependent angular radii \citep{Grandis2024}. Since these radii are uniquely determined by the cluster redshift, their explicit values are not stored. The inference network is therefore expected to learn the relation between the six shear measurements and the cluster redshift without requiring the radii as additional inputs.

Within the SBI training data generated for this work, the cluster member contamination fraction in the shear profiles, $f_\mathrm{cl}$, is neglected in the mock generator. $f_\mathrm{cl}$ depends on cluster richness, which is not included in the mock generation framework, and can instead be corrected for directly in the observational data, where the cluster richness is naturally available.
For the best-fit parameters of the eRASS1 cosmology, the generation of the shear profiles is performed in $t_\mathrm{shear} \approx 10s$.

\subsection{Performance and data structure}
\label{sect:performance}

For each simulation suite, one global cache of the eRASS1 background sky map, the eRASS1 maximum redshift map and the eRASS1 exposure map is created, which takes $t_\mathrm{map-cache} \approx 32$~s. The generation of one mock catalog on one CPU takes $t_\mathrm{mock} \approx 27$~s for a parameter set around the eRASS1 best-fit cosmology. 

We sample parameters from a uniform eleven-dimensional prior within the bounds listed in Tab \ref{tab:paramlist}. Mock catalogs with large $\Omega_\mathrm{m}$ and $\sigma_8$ or overall higher scaling relations (e.g., high $A_\mathrm{X}$) can include $\sim 10^7$ galaxy clusters and take more than $\sim 900$~s to be generated, while catalogs in different parts of the parameter space might include zero clusters and are thus generated almost instantly. Clearly, mock catalogs with $\sim 10^7$ clusters are considered unphysical given the observed eRASS1 sky with $\sim 3,000$ clusters. We therefore chose to decrease the prior ranges relative to \citetalias{Ghirardini2024}, while keeping them reasonably large.

The mock catalog creation is fully parallelizable, and we created \ncattot mock catalogs with a consistent rate of 18,000 catalogs per hour on $8\times144=1152$ CPUs within $t_\mathrm{total} \approx 11$~h. Each cluster within a mock catalog is represented by eight values: count-rate $\log C_\mathrm{R}$ (float), redshift $z$ (float), six shear values $\Vec{g}_\mathrm{t}=(g_\mathrm{t;1}, g_\mathrm{t;2}, g_\mathrm{t;3}, g_\mathrm{t;4}, g_\mathrm{t;5}, g_\mathrm{t;6})$ in six radial angular bins. The data vector of one cluster can be written as

\begin{equation}
    \vec{d} = \left(\log C_\mathrm{R}, z, \vec{g}_t\right),
\end{equation}
\noindent leading to an overall catalog shape of $(N_\mathrm{clus}, 8)$, with $N_\mathrm{clus}$ the individual number of clusters in each catalog.

\begin{table}[t]
\caption{Parameters of the forward model with prior ranges for all used parameters. The list represents all parameters used in \citetalias{Ghirardini2024}. For the SBI analysis, we have simplified the model to only include the parameters listed in the right-most column (Unused parameters are denoted by a dash). $\mathcal{U}(a, b)$ denotes a uniform distribution between $a$ and $b$, and $\delta(0)$ denotes a Dirac distribution at zero.}
\label{tab:paramlist}
\centering
\renewcommand{\arraystretch}{1.15}

\begin{tabular}{l l l}
\hline
Parameter & & Prior for SBI \\[4pt]
\hline 
\multicolumn{3}{l}{\textbf{Cosmology}} \\
\hline
Matter density              & $\Omega_{\mathrm{m}}$ & $\mathcal{U}(0.15, 0.45)$ \\
Power spectrum amplitude    & $\sigma_8$            & $\mathcal{U}(0.65, 1.05)$ \\
Hubble constant             & $H_0$                 & $\mathcal{U}(66, 70)$ \\
Baryon density              & $\Omega_{\mathrm{b}}$ & $\mathcal{U}(0.04, 0.05)$ \\
Scalar spectral index       & $n_{\mathrm{s}}$      & $\mathcal{U}(0.94, 0.98)$ \\[4pt]

\hline
\multicolumn{3}{l}{\textbf{X-ray scaling relation}} \\
\hline
Normalization               & $A_{\mathrm{X}}$      & $\mathcal{U}(0.01, 2.2)$ \\
Mass slope                  & $B_{\mathrm{X}}$      & $\mathcal{U}(0.1, 3.0)$ \\
Redshift slope              & $F_{\mathrm{X}}$      & $\mathcal{U}(-3, 3)$ \\
Redshift-mass mixing term   & $G_{\mathrm{X}}$      & $\mathcal{U}(-3, 3)$ \\
Intrinsic scatter           & $\sigma_{\mathrm{X}}$ & $\mathcal{U}(0.01, 1.5)$ \\[4pt]

\hline
\multicolumn{3}{l}{\textbf{Optical scaling relation}} \\
\hline
Normalization               & $A_{\mathrm{opt}}$    & - \\
Mass slope                  & $B_{\mathrm{opt}}$    & - \\
Redshift slope              & $C_{\mathrm{opt}}$    & - \\
Redshift-mass mixing term   & $D_{\mathrm{opt}}$    & - \\
Intrinsic scatter           & $\sigma_{\mathrm{opt}}$ &-\\[4pt]

\hline
\multicolumn{3}{l}{\textbf{WL mass bias}} \\
\hline
Redshift dependence 1          & $A_{\mathrm{WL}}$ & $\delta(0)$ \\
Redshift dependence 2          & $B_{\mathrm{WL}}$ & $\delta(0)$ \\
Mass slope scatter             & $C_{\mathrm{WL}}$ & $\delta(0)$ \\
Redshift dependence of scatter & $D_{\mathrm{WL}}$ & $\delta(0)$ \\[4pt]

\hline
\multicolumn{3}{l}{\textbf{Cross-correlations}} \\
\hline
X-ray--optical correlation coefficient & $\rho_{\mathrm{X,opt}}$ & - \\
X-ray--WL correlation coefficient      & $\rho_{\mathrm{X,WL}}$  & $\mathcal{U}(-1, 1)$ \\[4pt]

\hline
\multicolumn{3}{l}{\textbf{Contamination}} \\
\hline
Background fluctuations fraction & $f_{\mathrm{NC}}$  & - \\
AGN fraction                     & $f_{\mathrm{AGN}}$ & - \\
\hline
\end{tabular}
\end{table}

\section{Architecture}
\label{sect:architecture}
\begin{figure*}[h!]
    \centering
    \includegraphics[width=0.95\textwidth]{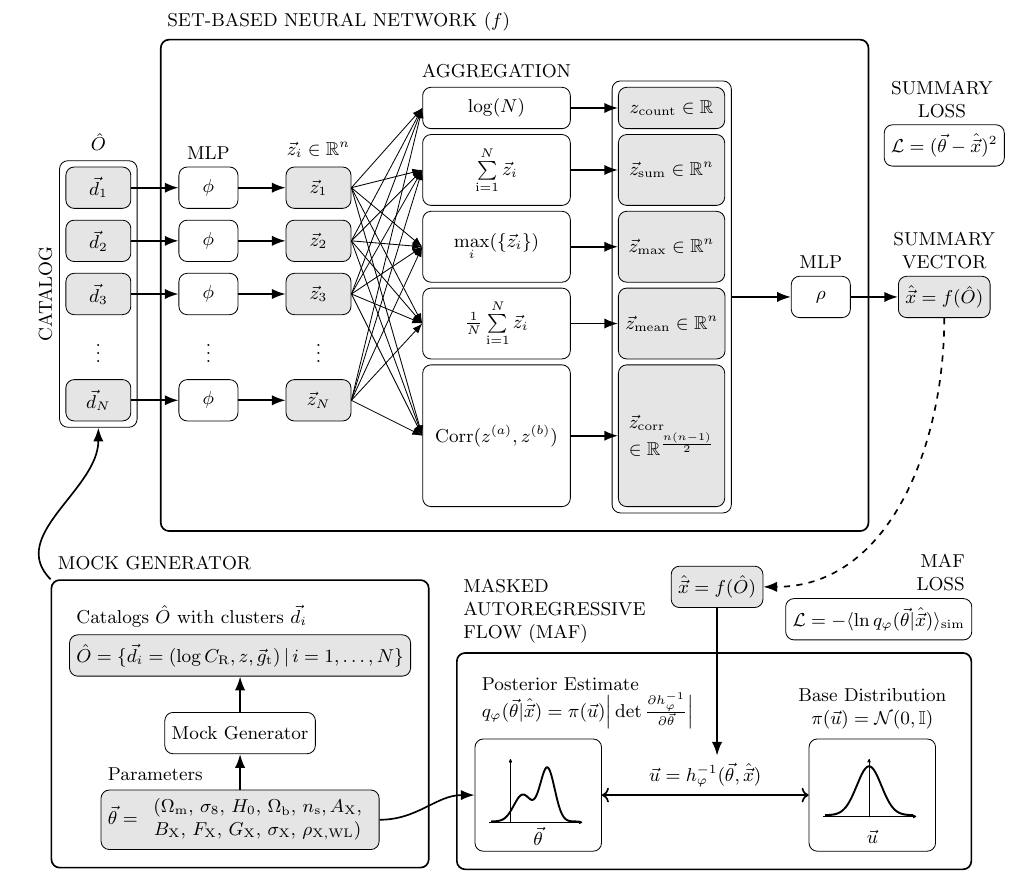}
    \caption{A two-stage neural architecture performs posterior estimation. First, a set-based network is trained to encode data catalogs into a fixed-length summary vector. Subsequently, a normalizing flow network is trained separately using these vectors to learn the posterior distribution.}
    \label{fig:architecture}
\end{figure*}

The neural network architecture used in this work has to match two requirements. First, it should take the complete catalog as an input and find a neural-informed summary statistic and second, it should yield Bayesian posterior estimates for the model parameters. 

In this work, cosmological inference is performed in two steps. A first set-based neural network summarizes the catalogs into highly informative summary statistics. These summary statistics serve two purposes: they compress the memory-intensive mock catalogs and convert them into a fixed shape independent of the varying input mock catalog shapes. In the second step, this summary is fed into an SBI pipeline that performs neural posterior inference over the desired parameters. 

This section describes the neural network architecture used to obtain neural-informed summary statistics. It describes the structure of the summary vector and how parameter inference is performed from there using SBI. A schematic overview of the two-fold neural architecture is presented in Fig.~\ref{fig:architecture}.

\subsection{Neural informed summary statistics}
\label{sect:perminvnn}

Traditionally, summary statistics are obtained using methods such as binning, correlation functions, bispectra, higher-order functions, or descriptive statistics. Depending on the expressiveness of the individual summaries, combinations might be used. In the context of galaxy cluster samples, binning is the most popular method \citep{Zubeldia2025}. The main challenge lies in finding a powerful binning scheme that represents data well over the full range of the parameter space. While a particular binning scheme might capture the data well in a specific region of the parameter space, it might do so poorly in other regions.

Alternatively, summary statistics can be obtained by leveraging state-of-the-art machine learning methods \citep[e.g.,][]{Fearnhead2010,Blum2012,Jiang2015,Charnock2018,Jeffrey2021,Lemos2023,Lehman2025}, which are employed in this work. Such a neural-informed summary statistic can be learned explicitly by training on the desired summary statistic, or implicitly by introducing a hidden layer in a neural network that represents a summary vector of the data (e.g., the last layer before the posterior estimation). Importantly, a summary statistic need not coincide with direct parameter predictions; it can instead be extracted from intermediate representations as a higher-dimensional and less restrictive encoding of the data, or be defined through alternative training objectives. Essentially, an optimized neural summary vector does not suffer from the usual limitations of rigid binning schemes, i.e. inconsistent bin widths. In contrast, a binning scheme might always include empty bins that carry very little information. Additionally, for the same reason, a neural-informed summary vector can be chosen to be larger than a binning scheme, as an increase in degrees of freedom is expected to facilitate information extraction more easily.

Finding a neural-informed summary vector using eRASS1-like galaxy cluster mock catalogs requires a summary network that handles varying numbers of objects arising from different parameter combinations, as well as stochastic processes such as the specific Poisson realizations of the number counts or the selection procedure. Catalog-like input objects exhibit a fundamental permutation-invariance symmetry: swapping any two clusters in the mock catalogs should not affect the summary vector. A class of architectures, called permutation-invariant or set-based neural networks, is designed to operate on set data, as in the case of the mock catalogs in this work.

We use a variation of Deep Sets~\citep{Zaheer2017}. In its original form, this architecture consists of two distinct neural networks, $\phi$ and $\rho$, each serving a distinct purpose. Written in a generalized form, Deep Sets are computed following the prescription

\begin{equation}
    f\left(\hat{O}\right) = \rho\,\left(\bigoplus_{i = 1}^N \phi(\vec{d}_i)\right).
\end{equation}

Here, $\hat{O}=\{\vec{d}_i=(\log C_\mathrm{R}, z, \vec{g}_\mathrm{t})_i|i=1,\dots,N\}$ is a full catalog consisting of $N$ elements (i.e., individual clusters) represented by $\vec{d}_i$. The $\bigoplus$ operation denotes an aggregation function. In the case of the original Deep Sets, outputs of the $\phi$ network are summed, and the result is given to $\rho$, hence $\bigoplus_\mathrm{DS} = \sum$. A similar architecture was used in the context of PointNet~\citep{Qi2016}, where the aggregation function is the maximum, hence $\bigoplus_\mathrm{PN}=\max$. 

A wide variety of different aggregation functions has been investigated since, including generalized means \citep{Kimura2024}, attention-based aggregators \citep{Chatzianastasis2022}, pairwise correlation aggregators for permutation-sensitive graph-neural networks (GNNs) \citep{Huang2022}, curve aggregators for PointNets \citep{Xiang2021}, and learnable aggregator functions for GNNs \citep{Ong2022}.

In this work, we use an aggregator built from a vector of permutation-invariant functions, including a cross-feature correlation function. The aggregator acts on $\mathcal{S}=\{\vec{z}_i | i=1,\dots,N\}$, denoting the set of outputs of the neural network $\phi$ (see Fig.~\ref{fig:architecture}). The cardinality of the set $N=|\mathcal{S}|$ equals the number of clusters. Each element of the set $\vec{z}_i \in \mathbb{R}^n$ is a feature vector with $n$ features. When referring to a specific feature $a$ in $\vec{z}_i$, we write $z_i^{(a)}$. The aggregator reads:

\begin{equation}
    \label{eq:aggregation}
    \bigoplus(\mathcal{S}) = 
    \begin{pmatrix}
        \log(N) \\
        \Sigma(\mathcal{S}) \\
        \mu(\mathcal{S}) \\
        \max(\mathcal{S}) \\
        \mathrm{vech}^+(\mathrm{Corr}(\mathcal{S}))
    \end{pmatrix},
\end{equation}

where $\log(N)$ is the logarithm of the cardinality of the set, and the functions $\Sigma, \mu$, and $\max$ are defined as 

\begin{align}
     \Sigma: &\, \mathbb{R}^{N\times n}\to\mathbb{R}^n; \quad z_i^{(a)} \mapsto \sum\limits_{i=1}^N z_i^{(a)}, \\
     \mu: &\, \mathbb{R}^{N\times n}\to\mathbb{R}^n; \quad z_i^{(a)} \mapsto \frac{1}{N} \sum\limits_{i=1}^N z_i^{(a)}, \\ 
     \max: &\, \mathbb{R}^{N\times n}\to\mathbb{R}^n; \quad z_i^{(a)} \mapsto \max\limits_i \left(\left\{z_i^{(a)}\right\}\right).
\end{align}

These functions thus each aggregate the $N$ outputs of $\phi$ with $n$ features each to a one-dimensional array with $n$ features.
The correlation aggregation refers to the empirical correlation between different features $a$ and $b$ of the output vectors $\vec{z}_i$. Hence it is a mapping $\mathrm{Corr}:\mathbb{R}^{N\times n}\to\mathbb{R}^{n\times n}$. Its matrix components $\mathrm{Corr}(\mathcal{S})_{ab}$ are defined as:

\begin{equation}
    \mathrm{Corr}(\mathcal{S})_{ab}
    =
    \frac{
    \sum_{i=1}^N \left(z_i^{(a)} - \mu^{(a)})(z_i^{(b)} - \mu^{(b)}\right)
    }{
    \sqrt{\sum_{i=1}^N \left(z_i^{(a)} - \mu^{(a)}\right)^2}
    \;\sqrt{\sum_{i=1}^N \left(z_i^{(b)} - \mu^{(b)}\right)^2}
    }, \\
\end{equation}

where $\mu^{(a)}$, $\mu^{(b)}$ are the mean values of the features along the cluster-dimension, defined by:

\begin{equation}
    \mu^{(k)} = \frac{1}{N} \sum_{i=1}^N z_i^{(k)}, \quad k\in\{a,b\}.
\end{equation}

The $\mathrm{vech}^+:\mathbb{R}^{n\times n}\to\mathbb{R}^{n(n-1)/2}$ operation flattens the upper right off-diagonal triangle of the correlation matrix to a one-dimensional array with dimension $n(n-1)/2$.

All of the listed aggregation functions preserve permutation invariance with respect to the ordering of the galaxy clusters in the catalog: Let $\vec{d}_i \in \hat{O}$ be the $i$-th cluster in the catalog. Then $\phi:\mathbb{R}^8\to\mathbb{R}^n$ maps every cluster $\vec{d}_i$ with 8 features on its latent representation $\vec{z}_i\in\mathcal{S}$ with $n$ features. Any two $\vec{z}_i, \vec{z}_j\in\mathcal{S}$ can be interchanged while preserving the result of any of the listed aggregation functions. Because the same function $\phi$ maps every cluster $\vec{d}_i$ to $\vec{z}_i$ for all $i$, the permutation invariance in the set $\mathcal{S}$ translates direclty to a permutation invariance of the clusters within a catalog $\hat{O}$. 

The dimension of the aggregation vector scales with the hyperparameter $n$ as

\begin{equation}
    \dim\!\left(\bigoplus(\mathcal{S})\right) 
    = 1 + n + n + n + \frac{n(n-1)}{2} 
    = \frac{n^2 + 5n}{2} + 1,
\end{equation}

which grows quadratically with $n$ and thus contributes to the architecture's overall memory requirements. This is mainly driven by the quadratic scaling of the correlation matrix. However, we found that including feature correlations is especially powerful, enabling meaningful constraints on the scaling relation parameters.

The aggregation functions $\mathrm{count}$, $\sum$, and $\max$ might introduce large output values in the aggregation layer. We therefore introduced a batch normalization layer after the aggregation to standardize potentially large values. We use a $\mathrm{ReLU}$ activation function in both networks, $\phi$ and $\rho$.

It has been shown that cluster counts cannot constrain the parameters $H_0$, $\Omega_\mathrm{b}$, $n_\mathrm{s}$, and $\rho_{X,WL}$ \citepalias[see, e.g.,][]{Ghirardini2024}. We therefore exclude those parameters from the loss function (i.e., we do not try to predict them) to avoid a high ground-truth loss due to noise in the parameters mentioned above. By constructing the loss function to include only well-constrained parameters, it is assured that the loss can be properly minimized. We emphasize that, nevertheless, all parameters listed in the right-most column in Tab.~\ref{tab:paramlist} are varied in the training data. This process of excluding the varied parameters from the loss function is equivalent to marginalizing over them. Given a summary network prediction $\hat{\vec{x}}~=~(\hat{\Omega}_\mathrm{m}, \hat{\sigma}_8, \hat{A}_\mathrm{X}, \hat{B}_\mathrm{X}, \hat{F}_\mathrm{X}, \hat{G}_\mathrm{X}, \hat{\sigma}_\mathrm{X})$ and a true label $\vec{\theta} = (\Omega_\mathrm{m}, \sigma_8, A_\mathrm{X}, B_\mathrm{X}, F_\mathrm{X}, G_\mathrm{X}, \sigma_\mathrm{X})$, we chose the following mean square error (MSE) loss function:

\begin{equation}
    \mathcal{L}_\mathrm{summary} = \mathrm{MSE}\left(\vec{\theta}, \hat{\vec{x}}\right) = \left(\hat{\vec{x}}-\vec{\theta}\right)^2.
\end{equation}

This neural network is trained to provide accurate proxies for the model parameters, which serve as the basis for subsequent posterior inference.

\subsection{Simulation-based inference}
\label{sect:sbiresults}
The highly informative summary statistic $\hat{\vec{x}}$ delivered by the set-based network (or alternatively an embedding vector) is used as input for Bayesian parameter inference. However, in general, there exists no model to construct a likelihood relating the abstract summary vector $\hat{\vec{x}}$ to the parameters $\vec{\theta}$ of interest. This setup, hence, requires a likelihood-free inference approach.

SBI enables Bayesian inference even when the likelihood is not tractable \citep{alsing_19,cranmer_20,papamakarios_17}. In SBI, inference is recast as a density-estimation problem. Given samples $(\vec{\theta},\hat{\vec{x}})$ from the joint probability density function (PDF) $p(\vec{\theta},\hat{\vec{x}})$, neural networks can be trained to either learn the density of simulations conditioned on the parameters $\mathcal{L}(\hat{\vec{x}}|\vec{\theta})$ (i.e. the likelihood) or conditioned on the data $p(\vec{\theta}|\hat{\vec{x}})$ (i.e. the posterior directly). 

Samples from the joint distribution are created through the mock generator described in Sect.~\ref{sect:trainingdata}. Initially, it generates a mock catalog $\hat{O}$ consisting of a stochastic, parameter-dependent number of clusters $\vec{d}$, each with 8 features. The resulting high-dimensional sample $(\vec{\theta}, \hat{O})$ is embedded into a lower-dimensional space through the set-based embedding network $f$, described in Sect.~\ref{sect:perminvnn}:

\begin{equation}
    \left(\vec{\theta}, \hat{\vec{x}}\right)\sim p(\vec{\theta}, \hat{\vec{x}}) \quad\mathrm{with}\quad \hat{\vec{x}}=f(\hat{O}).
\end{equation}

In principle, SBI could also be performed directly on samples $(\vec{\theta}, \hat{O})$ from the high-dimensional joint PDF. This is usually realized by attaching an embedding net in front of the normalizing flow and training it on the fly. In this work, however, we used a fixed embedding net to keep memory and training time requirements manageable.

While conditional density estimation can be approached traditionally, e.g., by considering histograms along the conditioning variable \citep[commonly known as approximate Bayesian computation][]{rubin_84, beaumont_02, akaret_15}, this approach breaks down in high dimensions. Modern implementations of SBI address this by using neural density estimators, often implemented as normalizing flows. For an in-depth explanation of this machine learning architecture for density estimation, we refer to~\citet{papamakarios_19}. In this work, we choose to learn the posterior directly \citep[an approach called neural posterior estimation][]{papamakarios_16,lueckmann_17}, therefore conditioning the density estimator on the summary vector $\hat{\vec{x}}$. 

To train the density estimator, a loss function is required. This loss function should minimize the discrepancy between the density estimate $q_\varphi(\vec{\theta}|\hat{\vec{x}})$ and the true posterior $p(\vec{\theta}|\hat{\vec{x}})$. Here, $\varphi$ are the neural network parameters. A natural choice for measuring this discrepancy between two distributions is the expected Kullback-Leibler (KL) divergence $D_\mathrm{KL}$ between the estimate and the posterior~\citep{kullback_51}. When used as a loss function, the KL divergence can be rewritten as posterior estimates averaged over a set of simulations~\citep{jordan_99,rezende_15,papamakarios_17}:

\begin{equation}
    \begin{aligned}
    \mathbb{E}_{p(\hat{\vec{x}})}\bigg[D_\mathrm{KL}&\Big(p( \vec{\theta}|\hat{\vec{x}}=f(\hat{O})\,||\,q_\varphi( \vec{\theta}|\hat{\vec{x}}=f(\hat{O})\Big)\bigg] \\
    &= \int d\hat{\vec{x}} p(\hat{\vec{x}}) \int d \vec{\theta} p( \vec{\theta}|\hat{\vec{x}}) \ln\left( \frac{p( \vec{\theta}|\hat{\vec{x}})}{q_\varphi ( \vec{\theta} | \hat{\vec{x}})} \right) \\
    &= \int d\vec{\theta} d\hat{\vec{x}} p(\vec{\theta},\hat{\vec{x}}) \ln \left( \frac{p( \vec{\theta}|\hat{\vec{x}})}{q_\varphi( \vec{\theta}|\hat{\vec{x}})} \right) \\
    &= -\mathbb{E}_{p( \vec{\theta}, \hat{\vec{x}})}[\ln q_\varphi( \vec{\theta}|\hat{\vec{x}})] + \mathrm{const.} \\
    &\approx -\frac{1}{N_{\mathrm{sim}}} \sum_n \ln q_\varphi( \vec{\theta}_n|\hat{\vec{x}}_n) + \mathrm{const.},
    \end{aligned}
\end{equation}

where $N_\mathrm{sim}$ is the number of simulations used for the evaluation. The aim of a training process is to minimize the loss. Therefore, the constant can be omitted, and the final loss for posterior estimation becomes

\begin{equation}
    \mathcal{L}=-\frac{1}{N_\mathrm{sim}} \sum_n \ln q_\varphi(\vec{\theta}_n|\hat{\vec{x}}_n).
\end{equation}

Note that this loss function is independent of the true posterior and, hence, does not require any explicit knowledge about it.

\section{Results and discussion}
\label{sect:results}

The neural networks described in Sect.~\ref{sect:architecture} have been trained on the mock data explained in Sect.~\ref{sect:trainingdata}. In this section, we describe the training process, the summary vector obtained, and the resulting posterior distributions. We have selected the model with the lowest validation loss that passed the calibration tests described in Sect.~\ref{sect:calibration}.

\subsection{Summary network}

The set-based neural network described in Sect.~\ref{sect:perminvnn} offers considerable flexibility in architectural design. In particular, the performance of the network is primarily influenced by five key components: the $\phi$ network operating on individual clusters, the choice of permutation-invariant aggregation function $\bigoplus$, the $\rho$ network acting at the population level, the method used for extracting a summary vector, and the selection of training hyperparameters.

\begin{figure}[h!]
    \centering
    \includegraphics[width=0.48\textwidth]{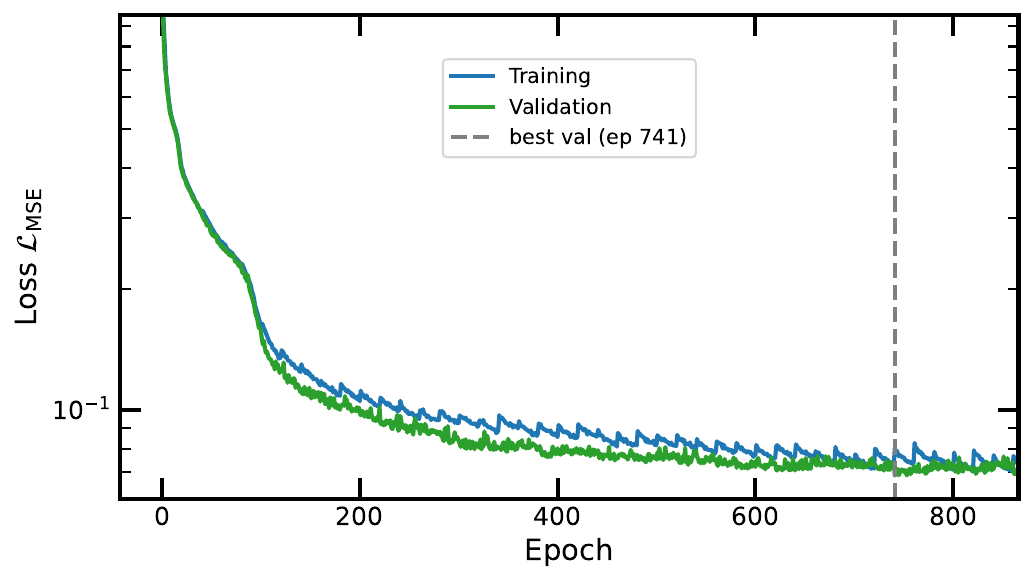}
    \caption{Loss curve of the set-based summary network. The blue curve corresponds to the MSE loss evaluated on the training data, and the orange one to the validation data. The spikes in the training loss curve originate from resampling \nepochtrain random catalogs from the total \ncattot catalogs every \epochsredraw epochs. One epoch thus corresponds to \nepochtrain random catalogs passed through the network once. The gray dashed line shows the epoch with the lowest validation loss, corresponding to the best network state.}
    \label{fig:regloss}
\end{figure}

For the two fully connected sub-networks $\phi$ and $\rho$, the freedom lies in the choice of the number of layers and nodes (hyperparameters) of the models. In the aggregation layer, the choice of the explicit aggregation functions is a hyperparameter. For the extraction of the summary vector, one can either take the parameter predictions from the last layer as described in Sect.~\ref{sect:perminvnn}, or take the output of one of the previous layers in the $\rho$ network. 
The results presented here have been obtained using the full aggregation vector from eq.~\ref{eq:aggregation}, and the parameter predictions.
The training hyperparameters are learning rate, batch size, and number of epochs until convergence.

We investigated the effect of changing the number of layers and nodes in $\rho$ and $\phi$. For $\phi$, the number of layers was varied from 2 to 5 and the nodes per layer from 16 to 256 (logarithmically spaced), while the architecture of $\rho$ was held constant. For $\rho$, the number of layers was varied from 2 to 10 while $\phi$ was held constant. Two types of layer structures were explored for $\rho$: a uniform structure with an equal number of nodes in all layers, and a funnel structure with a decreasing number of nodes in successive layers. Beyond these individual optimizations, we also tested several promising joint configurations of the $\rho$ and $\phi$ networks.
As expected, for the element-wise network $\phi$, we found that smaller networks starting from $\mathcal{O}(10^5)$ parameters are sufficient. The population-wise network $\rho$ should be chosen larger and deeper, with typically at least $\mathcal{O}(10^6)$ parameters. As described in Sect.~\ref{sect:perminvnn}, the last layer of the $\phi$ network can become a performance bottleneck (both, concerning memory and runtime) due to its $\mathcal{O}(n^2)$ scaling, where $n$ is the number of nodes in the last layer of $\phi$. We find that the network's overall performance is sensitive to the choice of $n$. From systematic tests in which we varied the number $n$ between 6 and 48, we conclude that $n \approx 30$ yields the best overall performance, while larger values tend to slow down learning, and smaller values reduce predictive power.

\begin{figure*}[h!]
    \centering
    \includegraphics[width=\textwidth]{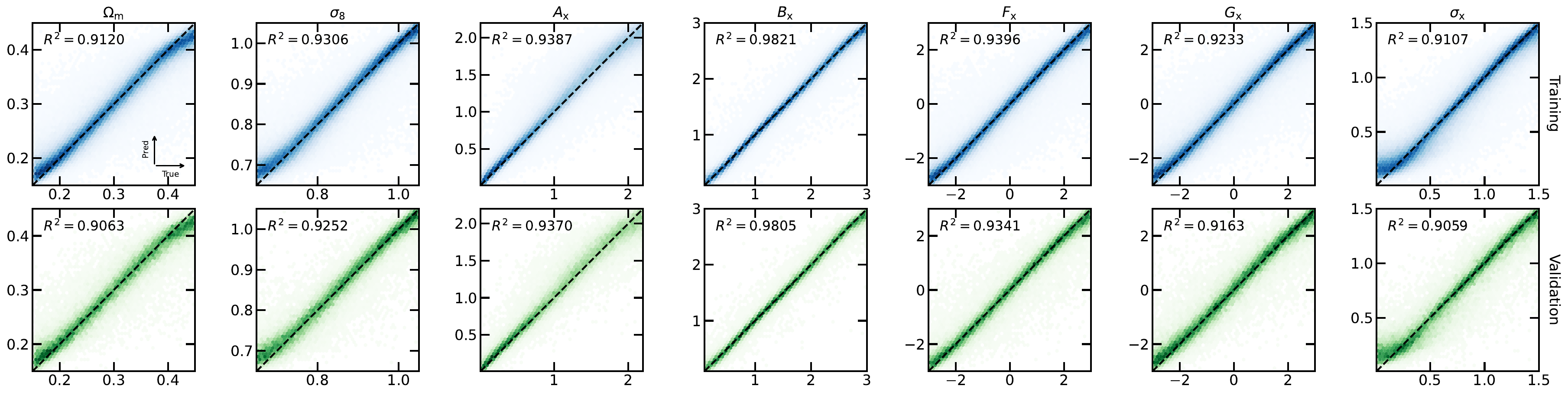}
    \caption{Binned parity plot of the constrained parameters. Each column represents one parameter with the training sample in the top tier and the validation sample in the bottom tier. The x-axis shows the ground truth, and the y-axis shows the network's prediction. The black-dashed line corresponds to a 1:1 relationship between truth and prediction. We expect the predictions to exhibit some scatter due to the stochastic processes underlying mock generation. The results are consistent across the training and validation sets, with only slightly larger $R^2$ values in the training set, indicating no overfitting. Even though the boundary regions show a slight trend toward the sample mean, the inference remains unaffected, as there is no bias in the region of interest.}
    \label{fig:parity}
\end{figure*}

For the aggregation layer, the most important function is the correlation matrix, while the count carries the least information, as expected. We found that removing the correlation matrix mostly reduces the neural network's constraining power on the scaling relation parameters, given in Tab.~\ref{tab:paramlist}. The sum $\sum$ and mean $\mu$ information, when combined, encode the count information, which is why removing the count aggregation from the aggregation functions has the smallest effect on the predictive power of the neural network.

For training and validation, a total of \ncattot mock catalogs have been created and split at a $9:1$ ratio. In the most optimal, non-compressed configuration, the total memory consumed by the training data exceeds \ramcattot. The actual memory requirements for training include the data itself, the neural network parameters, and the capacity to perform forward and backward passes. Memory demand typically peaks during backward passes. Consequently, even the largest available APUs (AMD Instinct MI300A APU) with 128~GB VRAM are insufficient for the desired task. To tackle this problem, we modified the training process. Instead of training on the full training sample at once, we draw a random subsample of \nepochtrain catalogs from the training data set and \nepochval from the validation data set and train the network on it for \epochsredraw epochs. Only then do we free the memory and reload another random set of mock catalogs. Note that one epoch does correspond to \nepochtrain randomly drawn training mock catalogs passed through the neural network once (instead of the full training data suite). As shown in Fig.~\ref{fig:regloss}, this process results in a loss function with systematic features. Every \epochsredraw epochs, the training loss naturally rapidly increases due to the updated training data. The validation loss tends to remain stable across changes in the data, indicating a stable learning process. Additionally, a second memory-optimization technique has been applied at the batch level by employing a subsampling method for unphysically large catalogs to reduce the input vector size. The details of this optimization are described in Appendix~\ref{app:aggregation}.

From the summary statistics, we find no significant difference in performance between using a highly compressed summary vector consisting of scalar parameter predictions, as described in Sect.~\ref{sect:perminvnn}, and employing a higher-dimensional previous layer (e.g., 512 dimensions). Consequently, the lower-dimensional parameter prediction representation is retained for simplicity. The resulting network performance for both the training and validation datasets is shown in Fig.~\ref{fig:parity}.

The best-performing models are obtained with learning rate $\eta=\reglearningrate$ and batch size $\texttt{bs}=\regbs$, with backward passes after each batch. Training is terminated once the validation loss of the summary network fails to improve for \regconverge consecutive epochs. Indepdently of the convergence patience, the network state with the best validation loss is used for all further analyses. Under these settings, the typical training time of the summary network was approximately $t_\mathrm{train}^\mathrm{sum}\approx \regtraintime$. All training runs are conducted on a hardware configuration with an Intel Xeon Ice Lake Platinum 8360Y CPU with 72 cores and 256 GB of memory, an Nvidia A100 GPU with 40 GB of VRAM. Among the available system configurations, this setup proved most effective for enabling efficient, low-latency data transfer between the disk, CPU, and GPU, which are the primary performance bottlenecks of the task.

\begin{figure}[h!]
    \centering
    \includegraphics[width=0.45\textwidth]{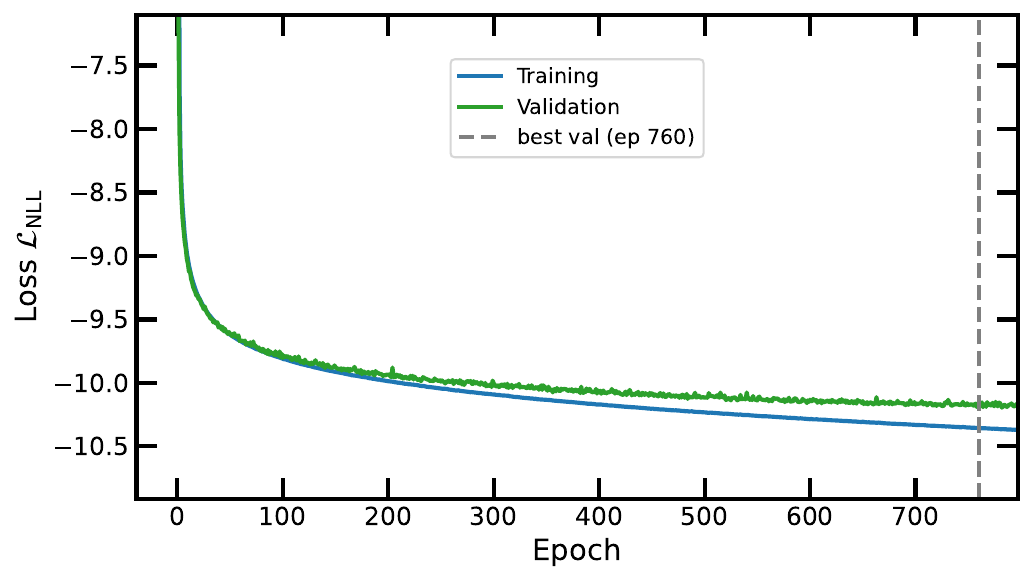}
    \caption{Training (blue) and validation (orange) loss curve of the SBI posterior estimation network. The gray dashed line indicates the lowest validation loss during the training process and corresponds to the best-performing model state.}
    \label{fig:sbiloss}
\end{figure}
\begin{figure*}[h!]
    \centering
    \includegraphics[width=0.95\textwidth]{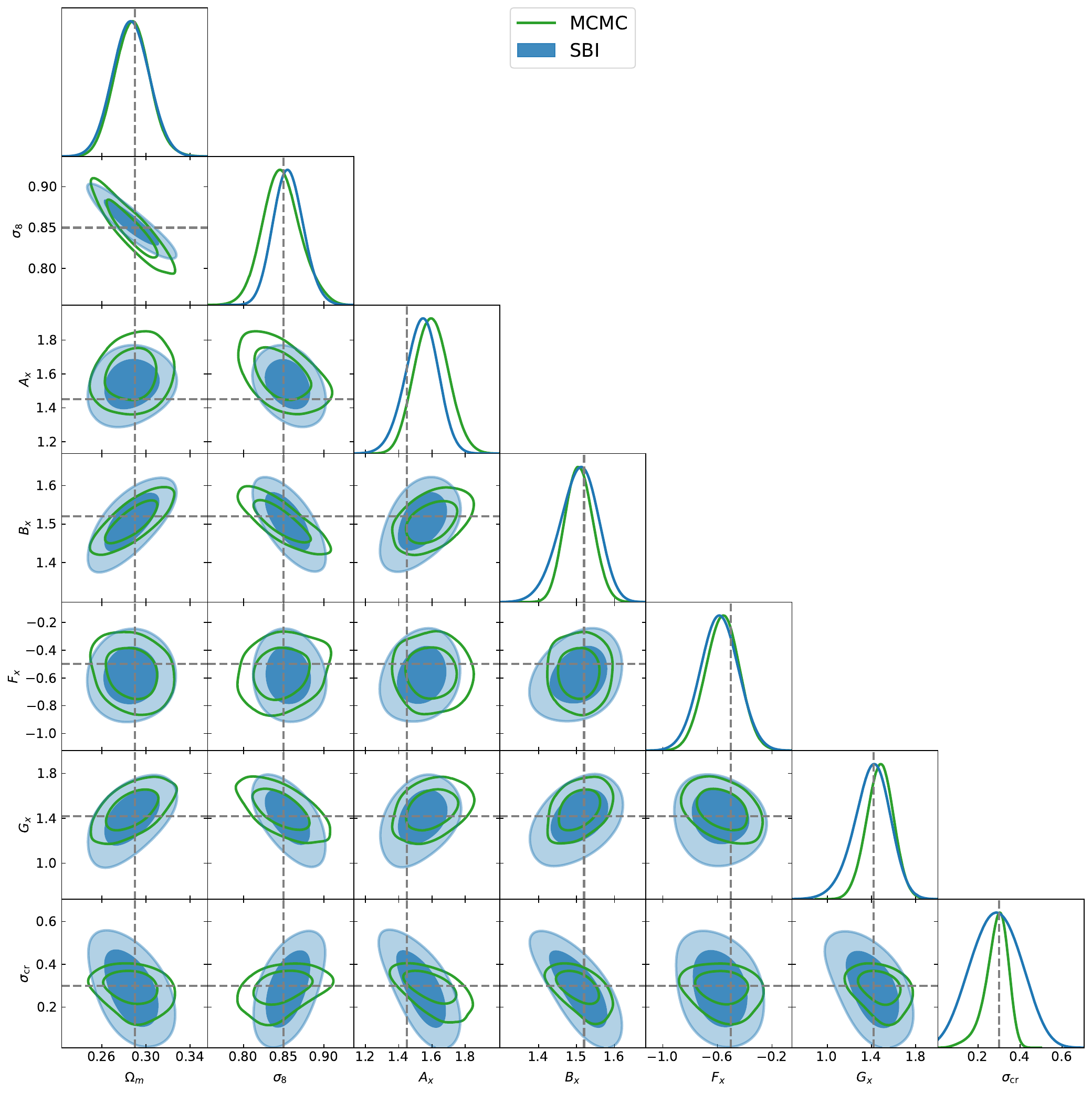}
    \caption{Example of a posterior projection plot obtained from inference on a representative mock catalog around the eRASS1 ($\extlike>10$) best fit values. The displayed posterior has uncertainties that closely match the mean uncertainties accross a suite of 64 mock catalogs around the fiducial parameter combinations. The dashed lines indicate the fiducial values.}
    \label{fig:corner}
\end{figure*}

We evaluate the network's performance using the coefficient of determination $R^2$, which measures the proportion of variance explained by the model. The metric provides a normalized goodness-of-fit that can serve as a scale-independent measure for comparing the network's performance across different output channels. Overall, when evaluated on the training data set, the goodness-of-fit is $R^2_\mathrm{train}=\rsqtrain$. On the validation data, it is $R^2_\mathrm{val}=\rsqval$. The parameter space can be split into quantities of explicit interest (cosmology, $\Omega_\mathrm{m}$ and $\sigma_8$) and auxiliary quantities (scaling relation parameters, $A_\mathrm{X}$, $B_\mathrm{X}$, $F_\mathrm{X}$, $G_\mathrm{X}$, $\sigma_\mathrm{X}$). For the matter density $\Omega_\mathrm{m}$, we find $R^2_{\mathrm{train}}(\Omega_\mathrm{m})=\rsqtrainomegam$ and $R^2_\mathrm{val}(\Omega_\mathrm{m})=\rsqvalomegam$, for the normalization of the matter power spectrum $\sigma_8$, we find $R^2_{\mathrm{train}}(\sigma_8)=\rsqtrainsigmaeight$ and $R^2_{\mathrm{val}}(\sigma_8)=\rsqvalsigmaeight$. Figure~\ref{fig:parity} visualizes the performance in a parity plot as well as the respective $R^2$ values in all output channels. 
After convergence, all catalogs are compressed using the trained summary network and stored. The resulting summary data consists of parameter proxies (summary vector) $\hat{\vec{x}}$ as inputs and true parameters $\vec{\theta}$ as labels as described in Sect.~\ref{sect:perminvnn}, and is used for posterior estimation in the following step.

\subsection{Neural posterior estimation}
\label{sect:nperesults}
The neural posterior estimation is sensitive to the choice of architecture and training hyperparameters. Furthermore, we find that the dimensionality of the target space (number of physical model parameters) can be a significant limiting factor for the performance of the posterior estimation with the presented method. The architecture consists of a concatenation of flow-based transformations, each with a specified number of layers. We achieved the best performance with a configuration with \sbimades flow-based transformations and \sbitransforms layers per transformation. Smaller networks tend to oversimplify features in the multidimensional posterior, i.e., the estimated posterior resembles a multivariate Gaussian. Larger networks did not improve the quality of the results but increased the training time.
Training and validation loss are displayed in Fig.~\ref{fig:sbiloss}.

The neural posterior estimator is trained with learning rates around $\eta = \sbilearningrate$ and batch sizes around $\texttt{bs}=\sbibs$. We stop the training when the validation loss of the SBI network does not improve for \sbiconverge epochs.
%In the model used for the main scope of this paper, we remove richness and the corresponding scaling relation. Furthermore, we apply a higher $\extlike>10$ cut than in \cite{Ghirardini2024} to achieve higher purity and reduce the impact of potential contaminants, as recommended by \cite{Bulbul2024}. we furthermore neglected the uncertainty on the weak lensing mass bias.
Owing to the modeling assumptions and simplifications described in Sec.~\ref{sect:trainingdata}, the analysis is reduced to 11 free parameters, including the unconstrained parameters $H_0$, $n_\mathrm{s}$, $\Omega_\mathrm{b}$, $\rho_\mathrm{X,WL}$). These parameters are represented by \ncattot\ mock catalogs sampled within the corresponding 11-dimensional prior volume (see Tab.~\ref{tab:paramlist}). The loss function is designed to take only the 7 constrainable parameters as input, with marginalization over the remaining unconstrained parameters. These results lead to the posterior shown in Fig.~\ref{fig:corner}.

The average uncertainties of the marginalized one-dimensional posteriors are reported in Tab.~\ref{tab:uncertainties}. These values are obtained by performing inference on 64 different mock catalog realizations around the fiducial cosmology and taking the arithmetic mean of the widths of the 68\% credible intervals of the marginalized posteriors.
The underlying mock data represents a subset of the eRASS1 cosmology sample; we expect it to be consistent with the cosmology analysis by \citepalias{Ghirardini2024} within the uncertainties.
We emphasize that removing WL mass bias uncertainties from the SBI training data in the modeling is a simplification, employed to enable better convergence of the SBI posteriors, that might underestimate systematic uncertainty relative to the more complete modeling in \citetalias{Ghirardini2024}. Overall, we find that the uncertainties on $\Omega_\mathrm{m}$ and $\sigma_8$ are \percentdeltaomegam\% and \percentdeltasigmaeight\%, respectively, for the full network constructed from 3,259 galaxy clusters. This level of precision is consistent with the cosmological parameter constraints obtained from the MCMC analysis of 5,259 clusters by \citetalias{Ghirardini2024}.

\begin{table}[]
    \centering
    \caption{Averaged 1$\sigma$ constraints on the model parameters, defined as the width of the marginalized 68\% credible intervals of the posterior distributions. The reported values are obtained by averaging over the uncertainties of the posteriors of 64 mock catalog realizations around the fiducial parameter combination.}
    \begin{tabular}{l r}
        \hline\hline
        Parameter & $1\sigma$--width \\
        \hline
        $\Delta\Omega_\mathrm{m}$ & \deltaomegam \\
        $\Delta\sigma_8$ & \deltasigmaeight \\
        $\Delta A_\mathrm{X}$ & \deltaAx \\
        $\Delta B_\mathrm{X}$ & \deltaBx \\
        $\Delta F_\mathrm{X}$ & \deltaFx \\
        $\Delta G_\mathrm{X}$ & \deltaGx \\
        $\Delta \sigma_\mathrm{X}$ & \deltasigmax \\
        \hline\hline
    \end{tabular}
    \label{tab:uncertainties}
\end{table}

We also tried to train the neural network on a more complex model with a larger parameter set, $\vec{\theta}$ \citepalias[see][]{Ghirardini2024}. However, we encountered difficulties in constraining the full parameter space starting from $\dim(\vec{\theta})\sim13$ and more parameters. In those tests with higher model complexity, the cosmological parameters of interest ($\Omega_\mathrm{m}$ and $\sigma_8$) were typically constrained, while scaling relation parameters remained almost unconstrained. 

\subsection{Calibration diagnostics}
\label{sect:calibration}

The quality of the obtained posteriors was validated using simulation-based calibration \citep[SBC,][]{Talts2018}, which provides an intrinsic verification of the accuracy of a Bayesian neural posterior estimator. It leverages the fact that the prior $\pi(\vec{\theta})$ should be recovered when averaging the posterior $p(\vec{\theta}|\hat{\vec{x}})$ over the data given the likelihood $\mathcal{L}(\hat{\vec{x}}|\vec{\theta})$:

\begin{equation}
    \pi(\vec{\theta}) = \iint \mathrm{d}\tilde{\vec{\theta}}\,\mathrm{d}\hat{\vec{x}}\,p(\vec{\theta}|\hat{\vec{x}})\,\mathcal{L}(\hat{\vec{x}}|\tilde{\vec{\theta}})\,\pi(\vec{\tilde{\vec{\theta}}}).
\end{equation}

\begin{figure*}[h!]
    \centering
    \includegraphics[width=0.95\textwidth]{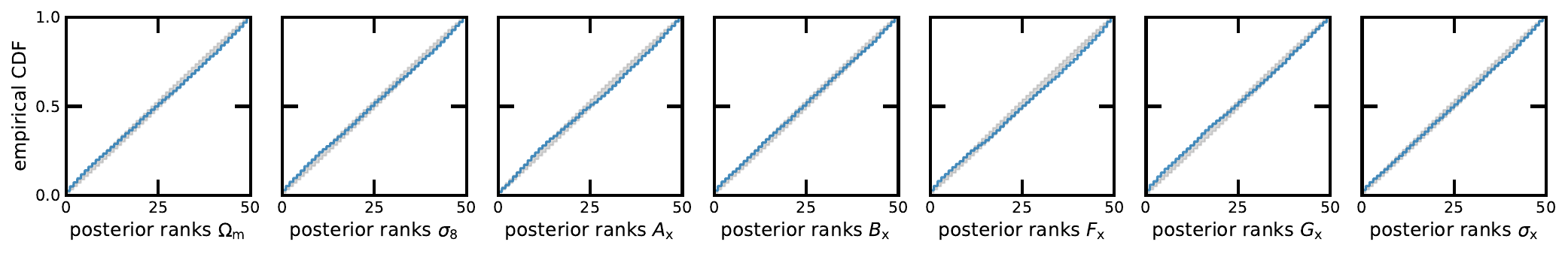}
    \caption{Model performance evaluation through an SBC rank plot. The blue lines indicate a cumulative probability density function of the rank statistic. The gray shaded areas represent the expected variance. The best model passes the SBC test with a slight tendency toward overconfidence.}
    \label{fig:sbc}
\end{figure*}

In practice, a sufficiently large sample $\tilde{\vec{\theta}}$ is drawn from the prior distribution $\pi(\tilde{\vec{\theta}})$. This prior sample is used to create mock catalogs $\hat{O}$. These mock catalogs are compressed into summary vectors $\hat{\vec{x}}$ through the set-based neural network. The pairs $(\tilde{\vec{\theta}}, \hat{\vec{x}}=f(\hat{{O})})$ implicitly contain the information on the unknown likelihood $\mathcal{L}(\hat{\vec{x}}|\tilde{\vec{\theta}})$. From the summary vectors $\hat{\vec{x}}$, the posterior $p(\vec{\theta}|\hat{\vec{x}})$, can be computed through the density estimation neural network. It can be shown that the rank of the original draw in a fixed number of samples from the corresponding posterior should be uniformly distributed, given an accurate posterior estimator \citep{Talts2018}. The cumulative distribution of the ranks for the trained posterior is shown in Fig.~\ref{fig:sbc}.
The posterior is accurate when the rank plots fall inside the gray error area for 99\% of the bins. A trend toward an S-shape indicates underconfidence, an inverted S-shape indicates overconfidence, and an inverted S-shape indicates a bias. We conclude that the presented neural posterior estimator is well calibrated with an almost unnoticeable trend toward overconfidence.

\begin{figure}[h!]
    \centering
    \includegraphics[width=0.5\textwidth]{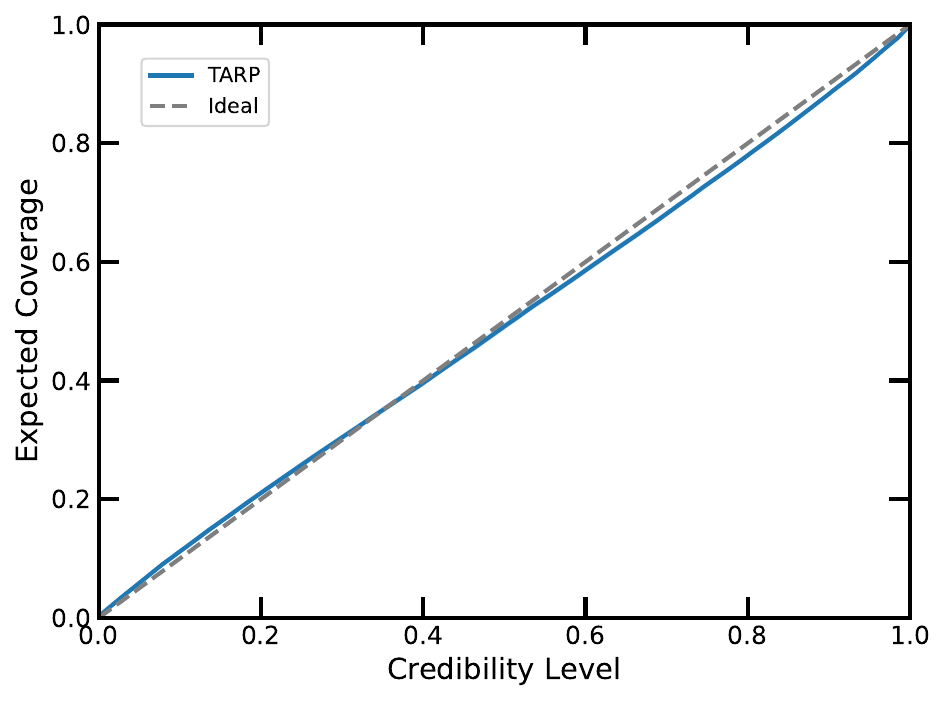}
    \caption{TARP diagnostic of the joint posterior distribution. The considered posterior has been marginalized over the scaling relation parameters and represents the posterior in $\Omega_\mathrm{m}$ and $\sigma_8$. The neural network passes the TARP coverage test.}
    \label{fig:tarp}
\end{figure}

To further assess the calibration of the posterior, we employ the Test of Accuracy with Random Points (TARP, \citealt{Lemos2023}). While SBC focuses on an evaluation of the marginalized posterior distributions, TARP can be used to validate joint posterior distributions. This is done by evaluating the coverage of credible regions around random reference points in the parameter space. Given an observation $\hat{\vec{x}}$ and its corresponding true parameter $\vec{\theta}^*$, TARP computes the distance from $\vec{\theta}^*$ to a set of random points $\{\vec{\theta}_\mathrm{r}\}$ and checks if $\vec{\theta}^*$ falls within the credible region defined by these distances relative to the posterior distribution $p(\vec{\theta}|\hat{\vec{x}})$. If the posterior estimator is well calibrated, the expected coverage probability should match the credibility level for all possible reference points. The TARP coverage curve provides a direct measure of the joint posterior accuracy, where deviations from the one-to-one line indicate miscalibration, such as under- or overconfidence, similar to the interpretation of rank statistics in SBC \citep[for further details, see][]{Lemos2023}. 
We compute the TARP diagnostic for the joint marginalized posterior of the parameters of interest, $\Omega_\mathrm{m}$, and $\sigma_8$. Figure~\ref{fig:tarp} shows good agreement between the expected coverage and the one-to-one line. We conclude that the neural posterior estimator passes the TARP test and is thus well calibrated. 

\subsection{Comparison with traditional MCMC Methods}
For comparison, we infer the same set of parameters using MCMC likelihood methods. To this end, we adopt the formalism developed by \citetalias{Ghirardini2024} as a baseline and adapt it to align with the modeling assumptions of this work. This section briefly summarizes these modifications; for a detailed description of the full likelihood computation, we refer the reader to \citetalias{Ghirardini2024}.

The Poisson likelihood introduced by \citetalias{Ghirardini2024} can be decomposed into four multidimensional integrals corresponding to the X-ray likelihood term, $\mathcal{L}_\mathrm{X-ray}$, the weak-lensing likelihood term, $\mathcal{L}_\mathrm{WL}$, the optical likelihood term, $\mathcal{L}_\mathrm{opt}$, and the prediction for the total number of clusters, $N_\mathrm{C}$, which serves as the Poisson parameter in the number-count analysis. The likelihood computation is embedded within a contamination model~\citepalias{Ghirardini2024}. We modify this formalism by completely omitting the optical likelihood term, $\mathcal{L}_\mathrm{opt}$, and marginalizing over the optical nuisance parameters and observed optical richness in the cluster number prediction, $N_\mathrm{C}$, consistent with the SBI modeling assumptions and motivated by the high sample purity of 97\% reported by \cite{Bulbul2024} for the simulated eRASS1 subsample, we do not include the contamination model. Cluster-member contamination of the tangential shear profiles is likewise neglected, consistent with the model described in Sect.~\ref{sect:shears}. For the selection function, we use a variant imposing an $\extlike > 10$ cut. The priors are adopted from Tab.~\ref{tab:paramlist}. We expect the MCMC pipeline to closely reproduce the mock-generation process. A direct comparison between the SBI and MCMC posteriors shows that both methods yield consistent results (Fig.~\ref{fig:corner}), with the fiducial parameters well recovered in both cases. The SBI framework provides tighter recovery of the intrinsic scatter in the X-ray scaling relation than the MCMC approach, which may be related to differences in the inferred constraints on the scaling parameters $A_\mathrm{X}$ and $F_\mathrm{X}$.

\begin{figure}[h!]
    \centering
    \includegraphics[width=0.5\textwidth]{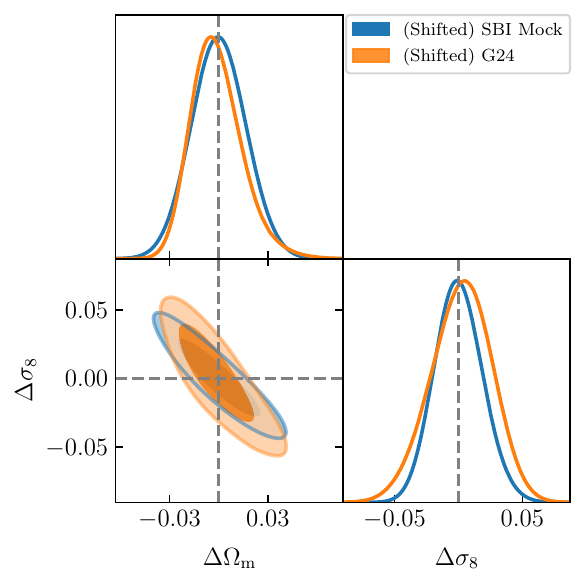}
    \caption{Comparison of cosmological posterior uncertainties from this work (a mock catalog with $\sim3{,}259$ clusters due to an $\mathcal{L}_\mathrm{ext}>10$ cut) and \citetalias{Ghirardini2024} (5,259 clusters due to an $\mathcal{L}_\mathrm{ext}>6$ cut). The posteriors from this work are inferred from a mock catalog that well represents the mock-averaged posterior uncertainty. The mock fiducial values can be chosen arbitrarily, so both contours have been shifted to their means for a direct comparison. The shift should not affect the interpretability of the comparison as the fiducial value matches the data-inferred value from \citetalias{Ghirardini2024} within $0.5\sigma$ in the $\Omega_\mathrm{m}$-$\sigma_8$ plane. Our analysis, based on realistic mock catalogs, recovers cosmological constraints with comparable precision to those obtained by \citetalias{Ghirardini2024}, despite using a subsample and omitting richness as an observable.}
    \label{fig:comparisonghirardini}
\end{figure}

As the simulated mock data mimic a higher $\extlike$-selection subsample of the eRASS1 galaxy cluster sample \citep{Bulbul2024}, it is instructive to compare the SBI posterior uncertainties with those obtained by \citetalias{Ghirardini2024}. While we expect the cosmological parameters to be consistent between the two analyses, the scaling-relation parameters may not be directly comparable. The higher $\extlike$ cut effectively restricts the sample to a subpopulation of more massive clusters with lower intrinsic scatter, potentially leading to different effective scaling relations. When comparing posteriors derived from data to those from mocks, it is important to note that the absolute position of the mock-inferred posterior is not uniquely defined and can be shifted to match the data-inferred posterior. We therefore shift both posteriors to their respective means and compare only their uncertainties. Figure~\ref{fig:comparisonghirardini} shows that, despite using mock catalogs derived from a smaller subsample of the eRASS1 catalog and not including richness as a mass proxy, the method is able to recover cosmological constraints with comparable precision under realistic mock conditions. This comparison should be interpreted as a validation of the methodology rather than a direct performance comparison on observational data. Overall, this highlights the robustness of the mock-based SBI approach in a controlled, yet realistic, setting.

\subsection{Related work}

Recent years have seen a growing number of applications of SBI methods to galaxy cluster catalogs, including eRASS1. Early attempts to constrain cosmology using galaxy cluster number counts within an SBI framework are presented by \cite{Reza2022} and \cite{Tam2022}. \cite{Reza2022} demonstrates the feasibility of deriving cosmological posteriors directly from simulations while simultaneously constructing efficient mock cluster catalogs. Their analysis employs true cluster masses from simulations without introducing a scaling relation between mass and observable. The number of clusters is binned into four mass bins across two redshift intervals, and selected using a step-function selection in true mass, $\Theta(M-10^{14}~\mathrm{M}_\odot)$. The resulting constraints depend sensitively on the strategy used to generate the training simulations. Specifically, the authors compared a set of 2,000 computationally expensive $N$-body simulations from the Quijote suite with a substantially larger sample of 11,378 computationally inexpensive realizations generated from an analytical halo mass function. The tightest constraints are obtained from the latter approach and are stronger than those derived in the present work. This outcome is not too surprising, given the idealized nature of their setup, which neglects the modeling of scaling relations and thereby avoids the additional uncertainties associated with a more realistic forward-modeling analysis and the approach we developed here.

An SBI approach tailored to X-ray cluster surveys is first proposed by \cite{Kosiba2025}. Their observables consisted of binned galaxy cluster distributions in a three-dimensional space of count rate, hardness ratio, and redshift, referred to as the X-ray observable diagram (XOD). The model adopts a step-function selection in count rate and filtered simulations by comparing their redshift distributions to a fiducial model, discarding parameter combinations identified as $>3\sigma$ outliers. The training set comprised 70,000 simulated XODs spanning eight model parameters. Cosmological posteriors for $\Omega_\mathrm{m}$ and $\sigma_8$ are inferred using Sequential Neural Posterior Estimation (SNPE), with the XODs compressed through a ResNet architecture and posteriors modeled using a Mixture Density Network.

Owing to their simplified modeling assumptions, the fiducial cosmology in \cite{Kosiba2025} yields approximately $4~\mathrm{clusters}~\mathrm{deg}^{-2}$, whereas the eRASS1 sample with $\extlike>10$ contains approximately $0.25~\mathrm{clusters}~\mathrm{deg}^{-2}$ and a total of 3,259 clusters \citep{Bulbul2024}. The statistical power of eRASS1 is therefore most closely matched by their $1,000~\mathrm{deg}^2$ realization containing roughly 4,000 clusters. Their resulting parameter uncertainties are approximately twice as large as those obtained in this work, likely reflecting the additional cosmological information provided here through the explicit inclusion of shear profiles. It should additionally be noted that \cite{Kosiba2025} implicitly marginalizes over nuisance parameters, whereas these are explicitly inferred in the present analysis.

Another relevant work by \cite{Regamy2026} introduced an SBI forward-modeling framework based on the distribution of cluster observables, namely X-ray flux, temperature, and redshift. The observables are summarized in a two-dimensional histogram of flux and temperature, supplemented by a one-dimensional histogram of redshift. Using Neural Posterior Estimation trained on 10,000 simulations, the authors constrained $\Omega_\mathrm{m}$, $\sigma_8$, and three free parameters describing the mass-luminosity scaling relation, while assuming an externally calibrated mass-temperature relation. Their selection function is implemented as a mass cut in true cluster mass. The analysis emphasized the importance of temperature information for breaking parameter degeneracies and showed that both $\Omega_\mathrm{m}$ and $\sigma_8$ are sensitive to the calibration of the mass--temperature relation, whereas the derived parameter $S_8$ remains comparatively robust. Several important methodological differences distinguish their framework from the present analysis. First, this work uses count rates rather than temperatures, since eROSITA cluster temperatures are associated with substantially larger observational uncertainties. Second, we adopt the eRASS1 selection function of \cite{Clerc2024}, calibrated on realistic simulated observations and defined directly in semi-observable space, rather than a cut in dark-matter mass, thereby providing a more realistic forward-modeling description. 
They consider a configuration with a cluster count comparable to eRASS1, achieved through increased survey depth rather than the shallow, wide-area survey strategy employed by eROSITA. Comparison with \citetalias{Ghirardini2024} revealed tighter constraints; however, these results rely on assumptions of precise eROSITA temperature measurements and a fully calibrated mass--temperature relation, conditions that are not presently realized in eROSITA-like surveys. Moreover, their model includes only six free parameters, compared to eleven in the present work, potentially leading to an underestimation of posterior uncertainties through the neglect of additional systematic effects.

An alternative framework, developed for a joint cluster and weak-lensing survey analysis, is introduced by \cite{Tam2022}. In addition to binned distributions of true cluster masses in redshift bins, the analysis incorporates binned shear profiles as observables. Their model further includes a selection process based on an observable mass proxy, $M'$, related to the true mass, $M$, through a lognormal intrinsic scatter of 10\%. The selection is implemented as a sharp, redshift-dependent mass threshold that accounts for the Eddington bias characteristic of observational surveys. Posterior constraints are presented for different background galaxy densities; however, even their lowest density of $20~\mathrm{gals}~\mathrm{arcmin}^{-2}$ exceeds the effective density of the DES survey used in this work by approximately a factor of 3.5 \citep{Gatti2021}. 

Despite methodological advances, direct comparisons across different SBI analyses remain challenging due to substantial differences in survey design, data dimensionality, and modeling choices \citep[e.g.][]{Zubeldia2025}. Variations in sky coverage, cluster number density, redshift distribution, and treatment of nuisance parameters can strongly affect both the precision and interpretability of cosmological constraints. Some studies achieve tighter posteriors by adopting idealized survey conditions or simplified scaling relations, whereas others incorporate more realistic observational effects but obtain broader uncertainties. This trade-off complicates any direct assessment of relative performance across methods.

Finally, it is worth noting that our methodology shares conceptual similarities with other works applying graph and set-based approaches to cosmological inference from galaxy catalogs. These studies include works by \citet{Anagnostidis2022}, \citet{VillanuevaDomingo2022}, \citet{Shao2023}, \citet{Makinen2022}, \citet{deSanti2023}, \citet{deSanti2025}, \citet{Massara2023}, and \citet{Roncoli2023}.

\section{Conclusions}
\label{sect:conclusion}
In this work, we present an application of implicit-likelihood inference for estimating cosmological parameters from variable-length catalogs based on eRASS1, an X-ray–selected galaxy cluster survey. To this end, we use two neural networks: one to generate a highly informative summary statistic and the other to estimate Bayesian posteriors from the summary vector. The networks are trained on highly realistic mock catalogs that include an X-ray scaling relation, the selection process, and the eROSITA instrument response. This work is therefore primarily focused on establishing and validating a methodological framework based on realistic simulations rather than deriving final cosmological constraints from observational data.

One aim of this study is to exploit the full information of the galaxy cluster mock catalogs. To achieve this, we avoid traditional compression methods such as binning and instead employ a neural network to construct a highly informative summary vector that reduces dataset complexity. Furthermore, the expressiveness of a binning scheme is highly dependent on the model parameters and is typically calibrated only on the data (i.e., for specific parameter values, the binning might be unfortunate, as mock data points lie, e.g., outside the grid or are clustered in a single bin). In practice, a varying number of galaxy clusters across training catalogs requires an architecture respecting the variable-length data structure. We therefore leverage set-based neural networks as embedding networks. By these means, we are able to fully compress the cluster mock catalogs down to proxies for the actual model parameters of interest, i.e., the summary vector $\hat{\vec{x}}$ consists of estimates for the cosmological parameters $\hat\Omega_\mathrm{m}$ and $\hat\sigma_8$ as well as the X-ray scaling relation parameters $\hat A_\mathrm{X}$, $\hat B_\mathrm{X}$, $\hat F_\mathrm{X}$, $\hat G_\mathrm{X}$, and $\hat\sigma_\mathrm{X}$, described in detail in Sect.~\ref{sect:trainingdata}.

Given this highly informative summary vector, a Bayesian neural density estimator is trained for the inference process. To this end, we use an MAF architecture that learns the posterior distribution. In practice, the MAF learns a function that transforms a simple base distribution, e.g., a normal distribution, into the posterior distribution, treating the input data as a condition. This input data is generated by compressing all mock catalogs using the permutation invariant network, thereby creating data-target pairs $(\hat{\vec{x}}, \vec{\theta})$, where $\vec{\theta}$ are the true model parameters (targets). Posterior samples are then created by drawing samples from the base distribution and transforming them into samples from the posterior through the trained MAF with the compressed catalogs $\hat{\vec{x}}$ as a condition. Using the training process with mock catalogs designed to resamble the eRASS1 sample with $\extlike > 10$ containing 3,259 galaxy clusters with a purity of 97\%, we obtain average uncertainties of \percentdeltaomegam\% on $\Omega_\mathrm{m}$ and \percentdeltasigmaeight\% on $\sigma_8$ over a suite of mock catalogs around the fiducial parameter values. We demonstrate that the method can recover constraints at a level comparable to those derived from MCMC analyses of 5,259 clusters by \citetalias{Ghirardini2024}. This validation is entirely based on highly realistic mock catalogs, ensuring that the performance assessment reflects practical survey conditions.

The power of the presented method lies in its great flexibility compared to MCMC methods. Evaluating likelihoods in MCMC methods amounts to solving computationally intensive integrals over grids of nuisance parameters (marginalization). These integrals are typically computed over a rigid box and can be sensitive to the choice of the number of grid points and integral bounds. Increasing the model complexity often introduces new nuisance parameters and potentially unfeasible increases in computation time and memory requirements proportional to the number of grid points in those nuisance parameters. While SBI brings its own challenges due to its recent and ongoing development (see Sect.~\ref{sect:perminvnn}), it does conceptually not struggle with increasing model complexity. The reason is that any change to the forward model (e.g., introducing a broken power-law scaling relation, adding a new mass proxy such as gas mass, or including additional systematic uncertainties) does not need to be explicitly modeled during training. It only has to be modeled implicitly in the mock simulator. Even if a more complex model is used in the mock simulator, the loss function does not have to include any new model parameters. Implicitly varying parameters in the training data that are not represented in the loss function correspond to marginalizing over those parameters (similar to the intrinsic quantities like dark matter mass or weak lensing mass of the mock clusters in the setup of this paper). 

As forward model complexity increases, the number of simulations used for training must increase, and so does the number of neural network parameters. In practice, generating simulations is highly parallelizable, and hence increasing the number of simulations is generally affordable. With the resampling method presented in this paper, the memory requirement of the training process is almost independent of the number of simulations, making the pipeline suitable even for an increasing number of simulations. An increase in the number of model parameters can also be compensated for by decreasing the number of subsampled simulations used for training during an epoch.

Ultimately, numerous improvements could be applied to the presented work. We present a possible list of suggestions for the future:
\begin{itemize}
    \item An increase in the number of nuisance parameters, as described above, does not compromise the applicability of the method. While additional nuisance parameters may reduce the constraining power, the framework remains applicable provided the increased complexity is properly accounted for, e.g., by using more mock data and more expressive networks. In this sense, the approach can accommodate a more complex mock generator that implicitly includes additional nuisance parameters. Similarly, it should be possible to extend the number of mass proxies for the clusters.
    \item An advanced approach would train the embedding net and the normalizing flow jointly, thereby eliminating the need to choose explictly construct a summary statistic. 
    \item The prior range could be extended and the posterior accuracy improved by using sequential neural posterior estimation \citep[SNPE, ][]{Wiqvist2021}. This technique involves training the network on a comparably small suite of simulations drawn from a broad prior distribution and then sequentially shrinking the prior range to the regions of interest and training on new simulations drawn from the smaller priors, while correcting for the bias introduced by changing the priors during inference. SNPE resembles state-of-the-art MCMC methods.
    \item The design of the aggregation layer offers a multitude of possibilities. It might be advantageous to use learnable aggregation functions. Furthermore, an attention-based transformer aggregator could further improve the network's performance if implemented in a memory-efficient way. 
\end{itemize}

While this work focuses on methodological validation using simulations, the ultimate aim is to perform inference using the eRASS1 data in follow-up work. To this end, several adaptations to the pipeline are required. In this analysis, we assume that all clusters possess DES-like shear profiles. However, in practice, the overlapping survey footprint of DES and eRASS1 covers only approximately 40\% of the eRASS1 sky area. However, in reality, the joint survey footprint of DES and eRASS1 covers only $\approx40\%$ of the eRASS1 sky area. Together with the DECADE survey, shear data can be obtained for $\approx 90\%$ of the observed clusters (Mistele et al. 2026, in prep.). The comparably small fraction of $\approx10\%$ clusters without a shear profile could be compensated by data injection methods (e.g., by assigning the average shear profile of similar clusters to clusters without a shear measurement). This has to be mirrored in the mock catalog generation and training process. Furthermore, we do not model the contamination of the shear profiles by cluster members in this paper, as this depends on cluster richness information that we omit in our simplified forward model. A possible workaround could be to correct for miscentering in the shear profiles at the data level, effectively resembling a partial backward modeling. Alternatively, the forward model could be extended to include richness, thereby accounting for shear miscentering. We expect that the presented SBI method works equally well with the data as the MCMC equivalent, given the consistency between the two methods on the mock data.

While we focus on eRASS1 galaxy cluster catalogs in this work, we emphasize the wide range of possible applications of the architecture. Any setup that includes a forward model producing variable-length catalogs can be trained with the presented combination of a set-based neural network and a normalizing flow. Especially, no explicit likelihood needs to be constructed; a realistic simulator is sufficient. Naturally, the input shape can be extended to images or even simulation boxes by adapting the set-based network accordingly.

In the next generation of cluster cosmology, survey data volumes are expected to increase by several orders of magnitude. The incorporation of multi-wavelength weak-lensing observations further increases the computational cost of constraining cosmological models within high-dimensional parameter spaces. In particular, analyses using forthcoming fourth-generation cosmology telescope data will require modeling approaches that are both computationally efficient and highly accurate, thereby motivating the development and use of simulation-based inference (SBI) methods. In this work, we present an SBI framework and demonstrate its performance using the eRASS1 dataset. The proposed methodology is readily extendable to upcoming large-scale optical surveys, including the Euclid and Nancy Grace Roman Space Telescopes \citep{Euclid2025,Han2023}. The true constraining power of this approach on observational survey data will be demonstrated in forthcoming work.

\begin{acknowledgement}
E. Bulbul, S. Zelmer, E. Artis, and X. Zhang acknowledge financial support from the European Research Council (ERC) Consolidator Grant under the European Union’s Horizon 2020 research and innovation program (grant agreement CoG DarkQuest No 101002585). K. Lehman acknowledges support via the KISS consortium (05D23WM1) funded by the German Federal Ministry of Education and Research BMBF in the ErUM-Data action plan. S. Krippendorf has been partially supported by STFC consolidated grants ST/T000694/1 and ST/X000664/1. N. Clerc acknowledges financial support from the Centre national d’études spatiales (CNES), France. N. Malavasi acknowledges funding by the European Union through a Marie Sk{\l}odowska-Curie Action Postdoctoral Fellowship (Grant Agreement: 101061448, project: MEMORY). Views and opinions expressed are however those of the author only and do not necessarily reflect those of the European Union or of the Research Executive Agency. Neither the European Union nor the granting authority can be held responsible for them.
\\

This work is based on data from \erosita, the soft X-ray instrument aboard SRG, a joint Russian-German science mission supported by the Russian Space Agency (Roskosmos), in the interests of the Russian Academy of Sciences represented by its Space Research Institute (IKI), and the Deutsches Zentrum f{\"{u}}r Luft und Raumfahrt (DLR). The SRG spacecraft was built by Lavochkin Association (NPOL) and its subcontractors and is operated by NPOL with support from the Max Planck Institute for Extraterrestrial Physics (MPE).
\\

The development and construction of the eROSITA X-ray instrument were led by MPE, with contributions from the Dr. Karl Remeis Observatory Bamberg \& ECAP (FAU Erlangen-Nuernberg), the University of Hamburg Observatory, the Leibniz Institute for Astrophysics Potsdam (AIP), and the Institute for Astronomy and Astrophysics of the University of T{\"{u}}bingen, with the support of DLR and the Max Planck Society. The Argelander Institute for Astronomy of the University of Bonn and the Ludwig Maximilians Universit{\"{a}}t Munich also participated in the science preparation for eROSITA.
\\

The eROSITA data shown here were processed using the eSASS/NRTA software system developed by the German eROSITA consortium.
\\

S. Zelmer thanks Nicole Hartman for an excellent introduction to the fundamentals of machine learning.
\\

The authors made limited use of large language models for minor language editing of individual sentences.
\\

\end{acknowledgement}

\bibliography{references}
\appendix

\section{Modeling of the miscentering}
\label{app:cvae}

Galaxy clusters in eRASS1 are detected by their X-ray emission. The peak of the X-ray emission defines the X-ray center, which is used to construct a tangential shear profile from optical data around this center. However, the X-ray center does not necessarily align with the center of mass of the cluster \citep{Grandis2024}. This affects the observed shear profiles and needs to be accounted for in the forward model.

A direct modeling of the miscentering would require the extent parameter \texttt{EXT} and the detection likelihood $\mathcal{L}_\mathrm{det}$ \citep{Grandis2024}, as computed by the eROSITA analysis software \texttt{eSASS} \citep{Brunner2018, Brunner2022, Merloni2024}. Since running \texttt{eSASS} on every mock catalog is computationally infeasible, we instead infer these quantities from observables available in the mock catalog using a conditional variational autoencoder (CVAE) trained on the eRASS1 digital twin by \cite{Seppi2022}.

We use the observables count rate $C_\mathrm{R}$, redshift $z$, exposure time $T_\mathrm{exp}$, and hydrogen column density $n_\mathrm{H}$ at the cluster position as conditioning variables. These are provided to both the encoder and decoder networks. The encoder maps the conditioning variables to a low-dimensional latent representation, while the decoder reconstructs \texttt{EXT} and $\mathcal{L}_\mathrm{det}$ from a sample drawn from the latent space together with the conditioning variables. Both encoder and decoder are implemented as fully connected neural networks.

We adopt a two-dimensional latent space, which we find sufficient to capture the variability in the simulated data. The network is trained by minimizing the standard CVAE loss, consisting of a reconstruction term and a Kullback–Leibler divergence regularization.

For each cluster, we draw from the learned latent distribution to generate samples of \texttt{EXT} and $\mathcal{L}_\mathrm{det}$, thereby capturing the intrinsic scatter in these quantities. The miscentering is then computed following the prescription of \cite{Grandis2024}.

\begin{figure}[h!]
\centering
\includegraphics[width=0.5\textwidth]{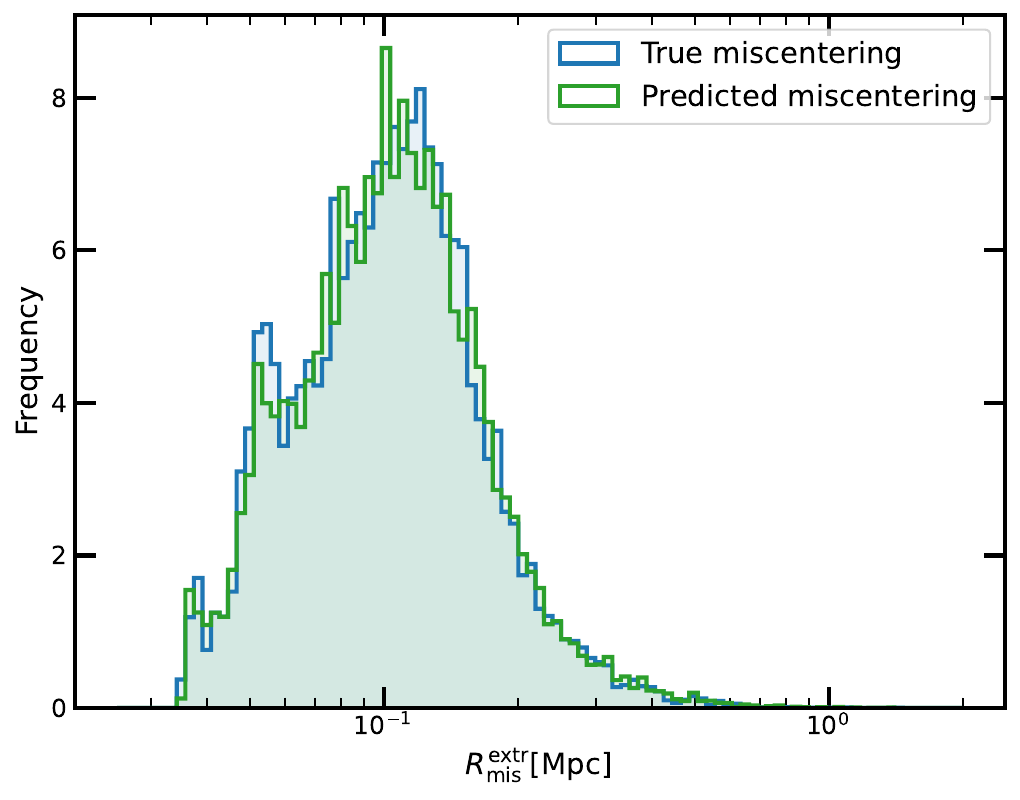}
\caption{Comparison of the miscentering distribution from the digital twin (blue line) and the prediction of the CVAE (green line), demonstrating good agreement.}
\label{fig}
\end{figure}

We split the simulated cluster sample into 80\% for training, 10\% for validation, and 10\% for testing. The resulting miscentering distribution is well reproduced by the CVAE, as shown in Fig.~\ref{fig}.

\section{Memory optimisation on batch level}
\label{app:aggregation}

The memory consumption is further determined by the size of individual catalogs. Even though, the set-based network is not dependent on the number of elements in the input set in theory, the technical implementation in $\texttt{pytorch}$ \citep{Paszke2019} requires a well-defined input tensor $T_\mathrm{in}$ for each batch with shape $(\texttt{bs},\, N_\mathrm{set},\, n_\mathrm{features})$. Here, $\texttt{bs}$ is the batch-size, $n_\mathrm{features}=8$ the number of features per object, and  $N_\mathrm{set}$ one global dimension for the number of objects per set. Since the number of clusters varies across one batch, the standard solution is to choose the number of clusters in the largest catalog as the global $N_\mathrm{set}$ and pad any empty entry in catalogs with less clusters with zero or not-a-number (NaN). However, given the underlying model, this leads to a large fraction of $\approx 90\%$ padded values on average. In terms of memory, this translates directly into a memory efficiency of $\approx 10\%$, i.e., only $\approx 10\%$ of the memory contains information. We solve this inefficiency by defining one maximum global $N_\mathrm{set}^\mathrm{max} = \regsubsample$. For each batch, we check, whether the largest catalog contains more than $N_\mathrm{set}^\mathrm{max}$ clusters and if so, we randomly sample \regsubsample clusters from it (while storing the original number of clusters $N_\mathrm{true}$). Those \regsubsample clusters are then inserted into the input tensor $T_\mathrm{in}$ for that batch together with the original number of clusters $N_\mathrm{true}$. 

\begin{figure*}[h!]
    \centering
    \includegraphics[width=\textwidth]{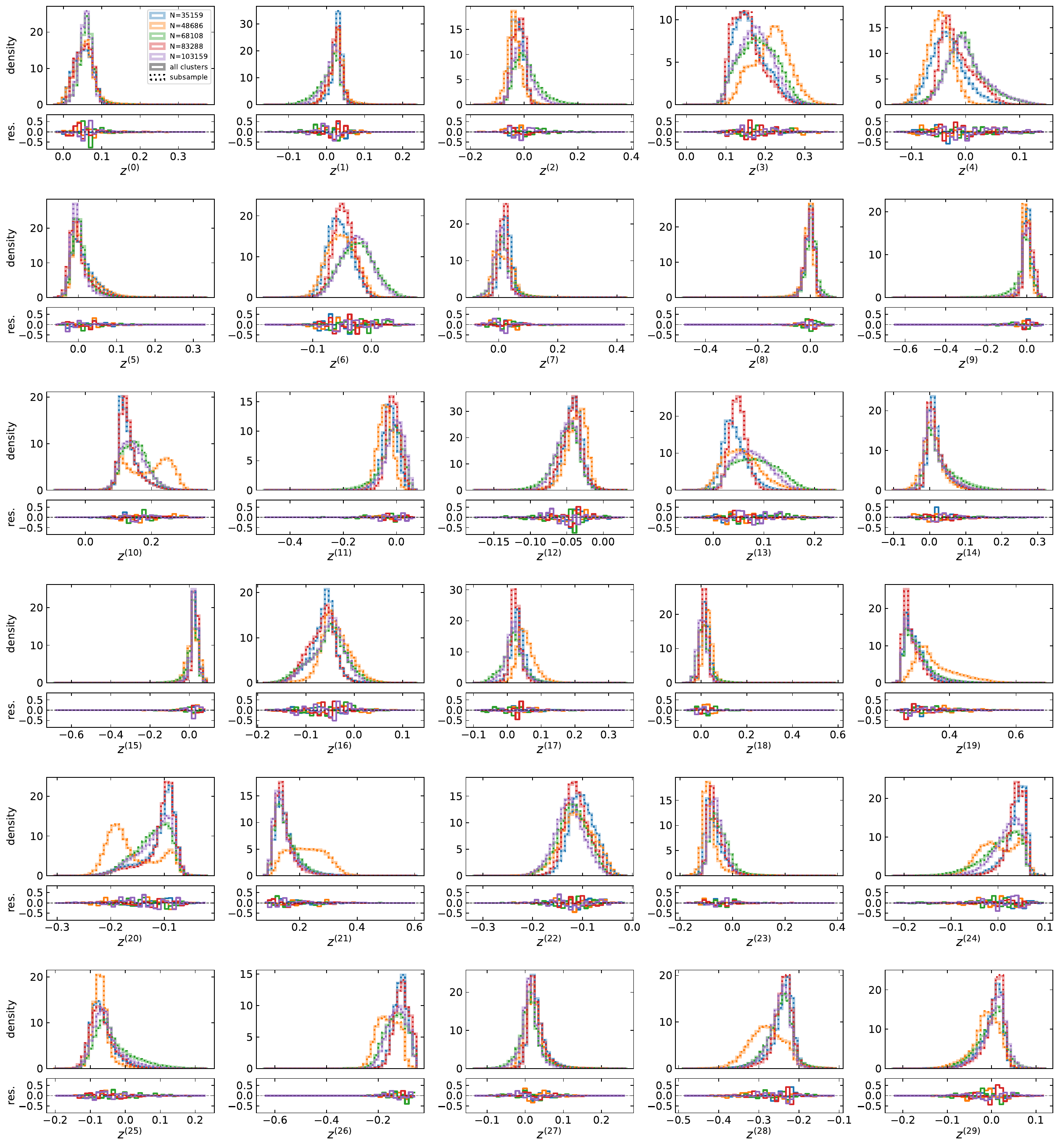}
    \caption{Output statistics of $z^{(a)}_i$ and residuals of the $\phi$ network for a subset of catalogs that required subsampling. No bias is introduced through the memory--saving sampling procedure applied to catalogs with more than \regsubsample clusters. The \regintermediatedim panels show normalized histograms of the output of the network layer right before aggregation for five representative mock catalogs with varying number of clusters (from $\sim~35,000$ up to $\sim~100,000$ clusters). The solid histograms show the output of the layer if all clusters have been used as an input, the dotted histograms show the same output if only a random subsample of \regsubsample clusters have been used as an input. For better visualization, the residuals between the histogram from the subsample and the full histogram are shown below each panel.}
    \label{fig:phinet}
\end{figure*}

The subsampling method described above requires corrections in the aggregation layer. The count needs to be passed explicitly for subsampled catalogs as it cannot be recovered from the output of the $\phi$ layer. We assume that statistically, the mean $\mu$ and the correlation of the $\phi$ output associated to a subsampled catalog is consistent with the one obtained from the full catalog. We furthermore assume that the sum $\sum$ of the $\phi$ outputs associated to subsampled inputs differs from the full catalog equivalent only by a factor $N_\mathrm{true}/N_\mathrm{set}^\mathrm{max}$, where $N_\mathrm{true}$ is the number of clusters in the full catalog, and $N_\mathrm{subsample}$ is the number of clusters in the subsample of the same catalog. We correct the aggregation function by this factor for the relevant catalogs. We note that the $\max$ aggregation function might get slightly biased low for subsampled catalogs due to selection effects. Exponential and power law tail fitting methods have been investigated to correct for the bias but led to unstable results. We decided to ignore the slight bias as we found no effect on posteriors for catalogs sampled in regions of the parameter space with cluster numbers around the observed eRASS1 catalog ($\approx 3000$). The effect of the subsampling on the last layer of the $\phi$ network, right before aggregation, is shown in Fig.~\ref{fig:phinet}.

\end{document}